\documentclass[
    prd, twocolumn, superscriptaddress, nofootinbib, amsmath, amssymb,
    aps, floatfix, preprintnumbers
]{revtex4-2}
\def\bath{{\cal B}}
\usepackage{style}
\usepackage{pifont}
\usepackage{aas_macros}

\begin{document}


\title{Dark Matter Freeze-In and Small-Scale Observables: \\ Novel Mass Bounds and Viable Particle Candidates}
\author{Francesco D'Eramo}
\email{francesco.deramo@pd.infn.it}
\affiliation{Dipartimento di Fisica e Astronomia, Universit\`a degli Studi di Padova, Via Marzolo 8, 35131 Padova, Italy}
\affiliation{Istituto Nazionale di Fisica Nucleare (INFN), Sezione di Padova, Via Marzolo 8, 35131 Padova, Italy}
\author{Alessandro Lenoci}
\email{alessandro.lenoci@mail.huji.ac.il}
\affiliation{Racah Institute of Physics, The Hebrew University, 91904, Jerusalem, Israel}
\affiliation{Laboratory for Elementary Particle Physics,
 Cornell University, Ithaca, NY 14853, USA}
\author{Ariane Dekker}
\email{ahdekker@uchicago.edu}
\affiliation{Kavli Institute for Cosmological Physics, University of Chicago, Chicago, IL 60637, USA}

\begin{abstract}
The suppression of cosmological structure at small scales is a key signature of dark matter (DM) produced via freeze-in in the low-mass regime. We present a comprehensive analysis of its impact, incorporating recent constraints from Milky Way satellite counts, strong gravitational lensing with JWST data, and the Lyman-$\alpha$ forest. We adopt a general strategy to translate existing warm dark matter (WDM) bounds into lower mass limits for a broad class of DM candidates characterized by quasi-thermal phase space distributions. The benefits of this approach include computational efficiency and the ability to explore a wide range of models. We derive model-independent bounds for DM produced via two-body decays, scatterings, and three-body decays, and apply the framework to concrete scenarios such as the Higgs portal, sterile neutrinos, axion-like particles, and the dark photon portal. Results from specific models confirm the validity of the model-independent analysis.
\end{abstract}

\maketitle

\section{Introduction}
\label{sec:intro}

Approximately half a century after its theoretical formulation, the Standard Model (SM) of particle physics continues to be confirmed by experiments at accelerators and colliders. Nevertheless, it leaves several fundamental questions about the universe unanswered. Chief among these is the nature of dark matter (DM), whose particle identity remains unknown~\cite{Jungman:1995df,Bertone:2004pz,Feng:2010gw,Salucci:2018hqu,Cirelli:2024ssz}. None of the numerous proposed theoretical candidates has yet received experimental confirmation.

In this paper, we focus on the \textit{freeze-in paradigm}, in which the hypothetical DM particle interacts so feebly with SM degrees of freedom that it never attains thermal equilibrium throughout cosmic history. Despite their feebleness, these interactions can still generate DM particles through occasional decays and scatterings involving constituents of the primordial thermal bath. Once produced, the DM particles redshift with the expansion of the universe and persist to the present day.

As is well known, when DM interactions with the bath are renormalizable, freeze-in production is most efficient when the temperature is comparable to the mass of the heaviest particle involved in the process~\cite{McDonald:2001vt,Kusenko:2006rh,Ibarra:2008kn,Hall:2009bx}. This implies that the DM relic abundance can be predicted solely in terms of particle physics parameters such as masses and couplings that could, in principle, be tested in laboratory experiments, without sensitivity to the unknown details of inflationary reheating. This feature is described by saying that freeze-in production is ``IR-dominated''. This is undoubtedly a strength of the framework. It is also important to recognize how the feeble interactions render it extremely challenging to test experimentally. Promising avenues  include displaced events at  colliders~\cite{Co:2015pka,Evans:2016zau,DEramo:2017ecx,Calibbi:2018fqf,Curtin:2018mvb,Belanger:2018sti,Junius:2019dci,No:2019gvl,Bae:2020dwf,Calibbi:2021fld}, direct detection~\cite{Chu:2011be,Essig:2015cda,Dvorkin:2019zdi,Boddy:2024vgt,Bernal:2024ndy}, and cosmological signatures imprinted on small-scale structure~\cite{Kamada:2013sh,McDonald:2015ljz,Roland:2016gli,Heeck:2017xbu,Bae:2017dpt,Boulebnane:2017fxw,Kamada:2019kpe,Dvorkin:2020xga,Ballesteros:2020adh,DEramo:2020gpr,Baumholzer:2020hvx,Egana-Ugrinovic:2021gnu,Du:2021jcj,Decant:2021mhj,Dienes:2021cxp,Xu:2024uas}.

The main subject of this paper is the latter effect. Freeze-in models can feature large free-streaming lengths, which erase perturbations below a characteristic scale and consequently suppress the formation of small-scale structure. It is therefore essential to employ probes that are sensitive to the DM distribution at these scales. Several observational strategies have been used to constrain DM properties. In particular, observations of ultra-faint dwarf galaxies, the Lyman-$\alpha$ forest, strong gravitational lensing, and stellar streams have all been successfully employed to place limits on models that suppress or enhance structure on small scales (see, e.g., the reviews provided by Refs.~\cite{Bullock:2017xww,Bechtol:2022koa} and the references cited in Sec.~\ref{sec:WDM}). It is important to recognize that these constraints are complementary, as they rely on different observational techniques and are affected by distinct systematic uncertainties. Our work examines the implications of this collection of small-scale probes.

The suppression of small-scale structure is a known consequence of Warm Dark Matter (WDM), whose cosmological effects have been extensively studied. This similarity with freeze-in DM allows us to construct a mapping between the two frameworks that provides multiple advantages. Most notably, it offers computational efficiency by avoiding the numerically intensive calculation of the matter power spectrum for each freeze-in model. In addition, it enables us to reinterpret existing mass bounds from the minimal WDM scenario. To implement this strategy, we solve the momentum-space Boltzmann equation to determine the DM phase space distribution (PSD) resulting from freeze-in production. We then constrain freeze-in DM by imposing limits on its \textit{warmness}, quantified by the second moment of the PSD.

We begin with a model-independent analysis of DM production via three different channels: two-body decays, binary scatterings, and three-body decays. Throughout, we assume that masses and couplings are chosen such that the DM relic abundance matches the observed value; to this end, we solve the relevant Boltzmann equation and compute the relic density self-consistently. Our general treatment assumes constant transition amplitudes, which is strictly valid only for two-body decays. In the case of scatterings and three-body decays, the matrix elements typically exhibit a nontrivial dependence on the momenta of the external particles. However, since freeze-in production is IR dominated and the relevant dynamics occur at a characteristic energy scale, this approximation remains well motivated. We confirm this expectation by analyzing several explicit microscopic realizations, where we compute the exact transition amplitudes and use them as input to the momentum-space Boltzmann equation.

The paper is structured as follows. In Sec.~\ref{sec:WDM}, we review the most up-to-date small-scale structure bounds on WDM and discuss the associated uncertainties and caveats. Section~\ref{sec:quasith} introduces the concept of a \textit{quasi-thermal} PSD and outlines a general strategy to translate WDM limits into bounds applicable to this broader class of candidates. In Sec.~\ref{sec:FI}, we focus on freeze-in production and derive model-independent lower bounds on the DM mass. Section~\ref{sec:UV1} presents explicit realizations of freeze-in via two-body decays in Higgs portal scenarios and in models involving sterile neutrinos. Section~\ref{sec:UV2} explores axion-like particles (ALPs) coupled to Standard Model fermions, while Sec.~\ref{sec:UV3} analyzes the dark photon portal. For each model, we compute the resulting PSD, extract the corresponding mass bounds, and verify consistency with the model-independent analysis of Sec.~\ref{sec:FI}. Technical details are deferred to the Appendices, and our conclusions are summarized in Sec.~\ref{sec:conclusions}. The main results of this work—both the model-independent analysis and the model-specific mass bounds—are compactly and clearly illustrated in the summary plot of Fig.~\ref{fig:models_summary}.

\section{Warm Dark Matter Mass Bounds}
\label{sec:WDM}
 
\begin{figure}
     \centering
\includegraphics[width=.9\linewidth]{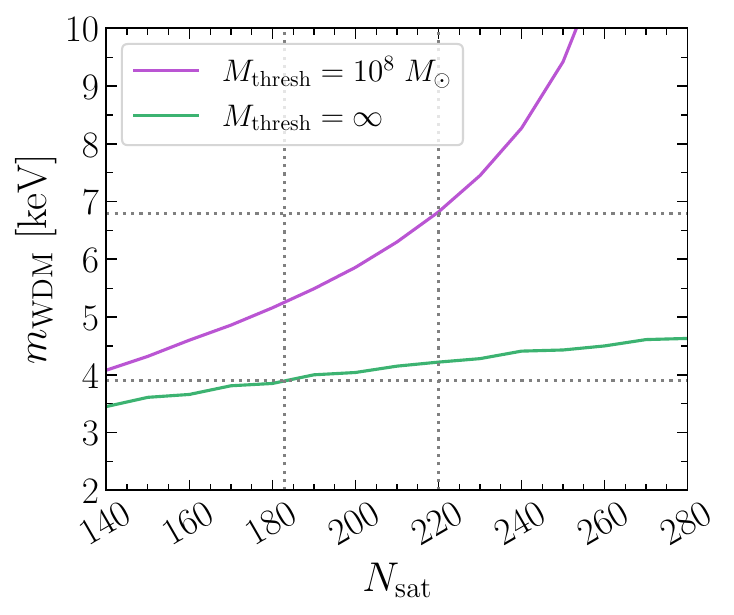}
     \caption{The constraints on $m_{\rm WDM}$ at 95\% CL, as a function of the number of observable satellite galaxies derived following the framework of Ref.~\cite{Dekker:2021scf}. The dotted lines show the limits adopted in this work. }
     \label{fig:Nsats}
 \end{figure}

\begin{figure}
    \centering
\includegraphics[width=1\linewidth]{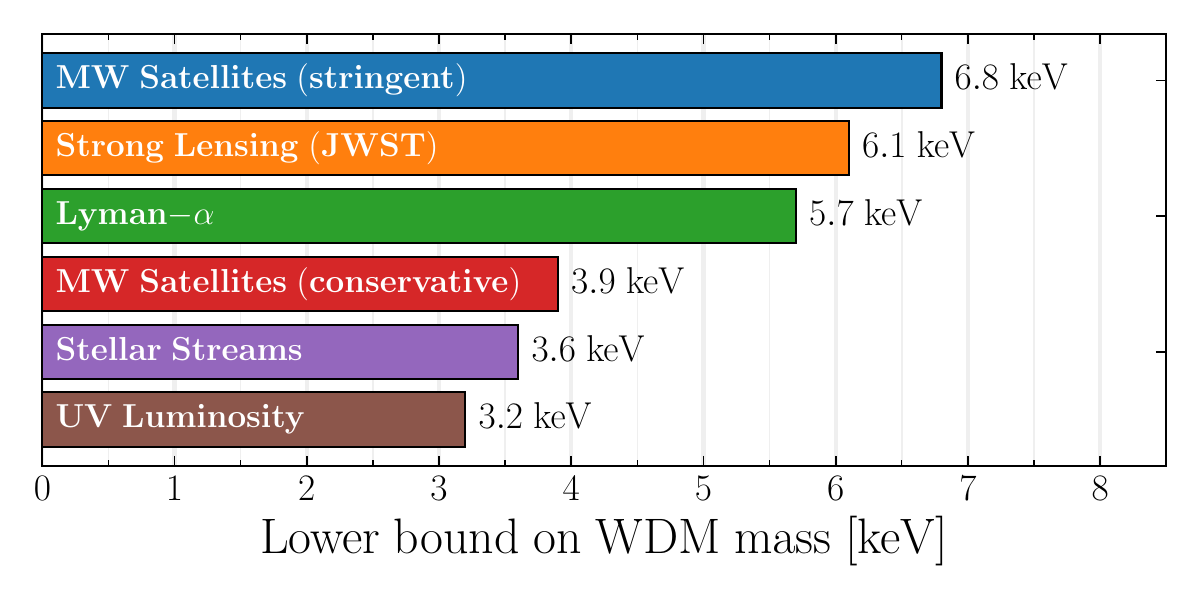}
    \caption{WDM mass bounds from small-scale observations: the \textcolor[HTML]{1f77b4} {\textbf{blue}} identifies the stringent bound from MW satellites; the \textcolor[HTML]{ff7f0f} {\textbf{orange}} from strong lensing due to JWST observations; the \textcolor[HTML]{2ba02b} {\textbf{green}} from Lyman-$\alpha$ forests; the \textcolor[HTML]{d62727} {\textbf{red}} the conservative bound from MW satellites; the \textcolor
   [HTML]{9467bd} {\textbf{violet}} from stellar streams; the \textcolor[HTML]{8c564c} {\textbf{brown}} from UV luminosity. See Sec.~\ref{sec:WDM} for details.}
    \label{fig:wdm-consraints}
\end{figure}

Small-scale tests of WDM are subject to significant observational and modeling uncertainties, making independent constraints from multiple probes essential. In this section, we review current WDM mass bounds, present an updated limit from Milky Way satellite counts with explicit assumptions (see Fig.~\ref{fig:Nsats}), and briefly discuss the prospects of combining probes in joint analyses. A summary of the mass bounds is shown in Fig.~\ref{fig:wdm-consraints}.

\textbf{Updated Milky Way satellite counts.} The number of satellite galaxies in the MW is expected to be reduced in the case of WDM, as the formation of low-mass subhalos capable of hosting galaxies is suppressed relative to CDM. Therefore, by comparing the predicted number of satellite galaxies in a given WDM model to the observed count ($N_{\rm obs}$), one can place strong constraints on WDM~\cite{Dekker:2021scf,DES:2020fxi,Newton:2020cog,Kennedy:2013uta,Lovell:2013ola,Schneider:2014rda,Polisensky:2010rw}. Uncertainties in these constraints arise, for instance, from the number of satellites associated with the Large Magellanic Cloud, the halo-galaxy connection~\cite{Bullock:2017xww}, and the mass of the MW halo, which is estimated to lie in the range $(0.5$--$1.5)\times 10^{12}\,M_{\odot}$~\cite{Cautun_2020,Callingham_2019,Eadie_2019,Posti_2019,Karukes_2020,Wang_2021,Bird_2022}.

Conservative limits can be obtained by assuming that all subhalos are capable of forming galaxies and by considering a wide range of MW halo masses. By combining data from the Dark Energy Survey (DES) and the Sloan Digital Sky Survey (SDSS) with $N_{\rm obs}=124^{+40}_{-27}$, Ref.~\cite{Newton:2020cog} finds $m_{\rm WDM} > 2.02$~keV (95\%~CL) after marginalizing over the MW halo mass. Moreover, using updated data from DES and Pan-STARRS1 with $N_{\rm obs} = 270$, Ref.~\cite{Dekker:2021scf} obtains conservative bounds of $m_{\rm WDM} > 3.6$--$5.1$~keV (95\%~CL), which vary only weakly with the MW halo mass. After correcting for anisotropic satellite distributions, $N_{\rm obs}$ was updated to $220 \pm 50$~\cite{DES:2019ltu}, which has only a minor impact on the conservative bounds. Incorporating a galaxy formation model leads to significantly stronger constraints, as it further reduces the total number of satellites. For instance, Ref.~\cite{DES:2020fxi} derives a bound of $m_{\rm WDM} > 6.5$~keV (95\%~CL) using an empirical galaxy--halo model and updated DES and Pan-STARRS1 data with $N_{\rm obs} = 220 \pm 50$.

When setting bounds using a galaxy--halo model, the uncertainty in the observed total number of satellites becomes important. In the CDM scenario, up to $\sim 150$ satellites are predicted to remain undetected, and approximately one-quarter of these are expected to be associated with the Large Magellanic Cloud (LMC)~\cite{DES:2019ltu}. Different studies adopt varying assumptions for $N_{\rm obs}$, leading to different constraints on the WDM mass.  Following the framework of Ref.~\cite{Dekker:2021scf}, we revisit WDM bounds for different values of $N_{\rm obs}$. We adopt a Milky Way halo mass of $M_{200}^{\rm DM} = 0.88 \times 10^{12}~M_{\odot}$, based on the weighted average of 20 individual estimates~\cite{2023ARep...67..812B}. Fig.~\ref{fig:Nsats} shows the derived 95\% CL limits on the WDM mass under two scenarios: a conservative case assuming all subhalos host galaxies (violet, $M_{\rm thresh} = \infty$), and a model with a galaxy formation threshold (green, $M_{\rm thresh} = 10^8~M_{\odot}$). As expected, the limits are more sensitive to $N_{\rm obs}$ when a galaxy formation threshold is applied, while in the conservative case, the dependence on $N_{\rm obs}$ is weak.  

Throughout this work, we adopt the galaxy formation threshold scenario with $N_{\rm obs} = 220$, which yields a lower bound of $m_{\rm WDM} > 6.8$~keV (95\% CL). For comparison, we also consider a conservative bound of $m_{\rm WDM} > 3.9$~keV (95\% CL), assuming all subhalos host galaxies and adopting $N_{\rm obs} = 183$, which excludes the estimated number of satellites associated with the LMC.

\textbf{Strong gravitational lensing.} DM subhalos within lens galaxies and halos along the line of sight can perturb strong gravitational lens images. These perturbations serve as probes of low-mass halos and allow constraints on WDM models based purely on the gravitational potential, thereby avoiding uncertainties related to baryonic physics. Systematic uncertainties arise from the modeling of the lens, source, and subhalo structures, as well as from flux ratio measurements. These are expected to improve with upcoming surveys and high-resolution follow-up observations~\cite{Nierenberg:2023tvi}. Surface brightness distortions from lensed quasar host galaxies, observed as arcs, have proven to be effective probes~\cite{Vegetti:2023mgp}. Using seven observations of multiply imaged quasars, Ref.~\cite{Hsueh:2019ynk} excludes $m_{\rm WDM} < 5.58$~keV at 95\% CL. Another method relies on the relative magnifications of lensed images~\cite{Gilman:2024mcs}. Based on Hubble Space Telescope observations of eight quadruply imaged quasars, Ref.~\cite{DES:2020fxi} reports $m_{\rm WDM} > 4.9$~keV at 95\% confidence. More recently, James Webb Space Telescope (JWST) observations of 31 lensed quasars strengthen this constraint to $m_{\rm WDM} > 6.1$~keV at posterior odds of 10:1~\cite{Keeley:2024brx}.

\textbf{Lyman-$\alpha$ forest.} The Lyman-$\alpha$ forest consists of absorption features in the spectra of high-redshift quasars, caused by neutral hydrogen in the intergalactic medium (IGM) along the line of sight. These absorption lines trace the underlying matter density field and can therefore be used to constrain WDM by comparing high-resolution spectral data with predictions from hydrodynamical simulations~\cite{Garzilli:2018jqh,Garzilli:2019qki,Baur:2015jsy}. These simulations carry significant uncertainties, particularly related to the thermal history of the IGM~\cite{Garzilli:2015iwa}. Using 1080 high-resolution cosmological hydrodynamical simulations along with quasar spectra from the Keck Observatory and the Very Large Telescope, Ref.~\cite{Villasenor:2022aiy} derives a lower limit of $m_{\rm WDM} > 3.1$~keV (95\% CL). More recently, Ref.~\cite{Irsic:2023equ} sets a stronger constraint of $m_{\rm WDM} > 5.7$~keV (95\% CL) using high-resolution spectra of distant quasars obtained with the HIRES and UVES spectrographs. This improvement is attributed to a larger dataset and reduced observational uncertainties, enabling the probing of smaller cosmological scales.

\textbf{Stellar streams.} DM subhalos that cross stellar tidal streams—formed by the disruption of globular clusters or dwarf galaxies—can induce density gaps and spurs along the stream. The number and size of these features provide insight into the subhalo mass function~\cite{Bovy:2016irg,Erkal:2015kqa}. Indeed, Refs.~\cite{Banik:2019smi,Banik:2019cza} find some of the tightest constraints on the WDM mass, deriving $m_{\rm WDM} > 3.6$~keV (95\% CL) by combining Gaia data with Pan-STARRS photometry of the GD-1 and Palomar 5 streams. Extracting DM subhalo properties from stream perturbations is, however, challenging due to degeneracies between the subhalo’s mass and its velocity~\cite{Ibata_2020,Bonaca:2024dgc}. Forthcoming data releases and improved numerical simulations are expected to reduce these uncertainties and enhance our ability to extract subhalo properties to test DM scenarios~\cite{S5:2024dcv,Carlberg:2024jne,LSSTDarkMatterGroup:2019mwo}.

\textbf{UV luminosity functions.} The UV luminosity function (UVLF), which measures the number density of galaxies as a function of UV magnitude, is a key probe of early galaxy formation. In WDM models, suppressed low-mass halo formation flattens the UVLF faint-end slope, enabling constraints from galaxy surveys at $z \gtrsim 4$ with HST and JWST~\cite{Corasaniti_2017,Schultz:2014eia,Menci_2016,Ellis:2025xju}. Recent JWST data yield the strongest bound to date, $m_{\rm WDM} \geq 3.2$~keV at 95\% CL~\cite{Liu:2024edl}. These constraints remain uncertain due to challenges in detecting high-redshift galaxies and modeling the halo mass–UV magnitude relation. Nevertheless, future JWST data will refine these limits and help disentangle baryonic effects~\cite{Maio:2022lzg,Ellis:2025xju}.

\textbf{Joint analyses.} Finally, joint constraints can be derived by combining different small-scale probes. For instance, Ref.~\cite{Enzi_2021} combines posterior distributions from strong gravitational lensing, the Lyman-$\alpha$ forest, and MW satellite counts, obtaining $m_{\rm WDM} > 6.048$~keV at 95\% CL by marginalizing over nuisance parameters. Ref.~\cite{Nadler_2021_} combines posteriors from strong lensing and satellite data, finding a Bayesian lower bound of $m_{\rm WDM} > 9.7$~keV at 95\% CL. Joint analyses require careful treatment of nuisance parameters, as shared constraints must use consistent parameterizations or be properly disentangled to avoid bias. For our purposes, we therefore focus on individual constraints.

\section{Setting mass bounds on quasi-thermal PSD: general strategy}
\label{sec:quasith}

The primary goal of this work is to track the PSD of freeze-in DM and derive the corresponding mass bounds. Before turning to the details of the freeze-in mechanism in the next section, we outline the general strategy.

Regardless of the production mechanism, DM generation in the early universe eventually becomes inefficient, after which the evolution of the PSD is governed solely by cosmological redshift. We focus on this asymptotic form of the PSD and quantify its \textit{warmness} through the second moment. To assess whether a given solution is viable, we compare its warmness to that of a thermal WDM particle whose mass saturates the observational bounds collected in Sec.~\ref{sec:WDM}. The minimum mass for freeze-in DM is then defined as the value that yields the same warmness. This approach to setting mass bounds is justified by the thermal origin of the produced DM particles: although they never reach thermal equilibrium, they inherit typical energies comparable to the bath temperature.

A more comprehensive analysis would employ the full PSD to evaluate the DM impact on the formation of cosmological structures. However, this approach is computationally expensive for deriving bounds in concrete models and impractical for a systematic exploration of their parameter space. Instead, we build on the results of Ref.~\cite{DEramo:2020gpr}, which established a connection between freeze-in DM and WDM. Using the Boltzmann solver \texttt{CLASS}~\cite{Lesgourgues:2011rh}, it was shown in a model-independent manner that the matter power spectrum for freeze-in DM exhibits features closely resembling those of WDM. In particular, the relative ``transfer function''\footnote{Following standard conventions, we define the squared transfer function as the ratio of the linear power spectrum for freeze-in DM to that of $\Lambda$CDM: $\mathcal{T}^2(k) \equiv P(k) / P_{\Lambda\rm CDM}(k)$.} displays a sharp cutoff at large wavenumbers, indicating a suppression of small-scale structure that mirrors the behavior of WDM. As a result, the freeze-in transfer function for a given model can be effectively mapped onto that of WDM by selecting an appropriate value of $m_{\rm WDM}$. Our strategy is thus supported qualitatively by the thermal origin of freeze-in DM and quantitatively by the results of Ref.~\cite{DEramo:2020gpr}.

In light of the above, freeze-in DM production is just one example where our strategy is effective. The same approach applies more broadly, provided two conditions are met. First, the DM PSD must be such that its physical effects are well captured by its first three moments.\footnote{The zeroth moment corresponds to the PSD normalization, fixed by the observed DM relic density.} Second, the first ($\langle p \rangle$) and second ($\langle p^2 \rangle$) moments must satisfy $\langle p \rangle^2 \sim \langle p^2 \rangle$. The freeze-in mechanism is a prototypical case where both conditions are naturally satisfied due to the thermal origin of the produced DM particles.

We refer to this class of DM candidates as having a \textit{quasi-thermal} PSD. For illustration purposes, we provide an analytical expression for a generic DM candidate $\chi$ that captures its typical behavior
\begin{equation}
\text{Quasi-thermal PSD:} \qquad f_\chi^{\rm qth}(q) \propto \frac{q^{\alpha}}{e^{\beta q} + \kappa} \ .
\label{eq:QTPSD}
\end{equation}
First, we note that the quasi-thermal PSD is defined up to an overall normalization fixed by the relic density. Second, the Friedmann-Robertson-Walker (FRW) geometry ensures that the PSD depends only on FRW time and the magnitude of the physical momentum. We express the PSD in terms of the comoving momentum $q$, formally defined in the next paragraph. Importantly, $q$ remains constant once DM production shuts off, naturally factoring out cosmological redshift. Consequently, the functional form in Eq.~\eqref{eq:QTPSD} describes the asymptotic, time-independent PSD in the late universe. The parameters $(\alpha, \beta, \kappa)$ are real, with $\kappa$ restricted to $\{-1, 0, +1\}$. For instance, WDM corresponds to a thermal PSD with $(\alpha, \beta) = (0, 1)$ and $\kappa$ determined by particle statistics. We require $\beta > 0$ for convergence of the normalization integral and $|\alpha| \lesssim 1$ to keep the PSD close to thermal. This is true for all freeze-in scenarios studied in this work. We stress once again that Eq.~\eqref{eq:QTPSD} is only for illustrative purposes and no results of this work depend on it.

We now return to the formal definition of comoving momentum, which will be essential in what follows. The spatial homogeneity and isotropy of the FRW geometry ensure that the PSD of the DM particle $\chi$ depends only on the FRW time $t$ and the magnitude of the physical momentum $p(t)$. Even after interactions cease and Hubble expansion dominates, $p(t)$ remains time-dependent due to cosmic redshift, scaling as $p(t) \propto a(t)^{-1}$, where $a(t)$ is the FRW scale factor. To factor out this dependence, we introduce the dimensionless comoving momentum
\begin{equation}
q \equiv \frac{p(t)}{T_P} \frac{a(t)}{a(t_P)} \ .
\label{eq:qcom}
\end{equation}
Here, $t_P$ is an \textit{arbitrary} reference time, and $T_P$ is the bath temperature at that time. While the choice of $t_P$ is conventional and does not affect physical results, a suitable selection can simplify the analysis. In the freeze-in scenarios considered in this work, where production is IR dominated and proceeds through decays or scatterings mediated by renormalizable interactions, a convenient choice for $T_P$ is the mass of the heaviest particle involved in the production process. Explicit examples of this choice are provided in the next section. As noted above, the PSD expressed in terms of the comoving momentum defined in Eq.~\eqref{eq:qcom} becomes time-independent once DM production has ceased.

The relation between $t_P$ and $T_P$ follows from the dynamics of the expansion. We assume a standard cosmological history where, at early times, the energy density is dominated by a hot plasma in thermal equilibrium. The scale factor evolves as $a(t) \propto t^{1/2}$, and the time-temperature relation is set by the Friedmann equation
\begin{equation}
H(t) = \frac{1}{2 t} = \frac{\pi \sqrt{g_\star(T)}}{3 \sqrt{10}}  \frac{T^2}{\Mpl} \ .
\label{eq:HRD}
\end{equation}
We adopt the results of Ref.~\cite{Laine:2015kra} for the effective number of relativistic degrees of freedom, $g_\star(T)$. Likewise, the entropy density is $s = (2 \pi^2 / 45)\, g_{\star s}(T)\, T^3$, where $g_{\star s}(T)$ is also taken from Ref.~\cite{Laine:2015kra}. 

Even if DM particles never reach thermal equilibrium, it is convenient to define an effective DM temperature
\begin{equation}
T_\chi = T_P \frac{a(T_P)}{a(T)} = T \left( \frac{g_{\star s}(T)}{g_{\star s}(T_P)} \right)^{1/3} \ ,
\end{equation}
where $a(T)$ is the FRW scale factor expressed as a function of the bath temperature $T$ via the relation in Eq.~\eqref{eq:HRD}. The first equality allows us to recast the comoving momentum as $q = p(t) / T_\chi$, while the second follows from entropy conservation ($s a^3 = \text{const}$). As pointed out by Refs.~\cite{Kamada:2013sh,McDonald:2015ljz,Roland:2016gli,Heeck:2017xbu,Kamada:2019kpe} and quantitatively confirmed by Ref.~\cite{DEramo:2020gpr}, a key quantity for cosmological structure formation is the DM root-mean-square velocity
\begin{equation}
W_\chi \equiv \frac{\sqrt{\langle p^2 \rangle}}{m_\chi} = \frac{T_\chi}{m_\chi} \, \sigma_q  \ ,
\label{eq:Wchi}
\end{equation}
where the dimensionless PSD second moment is 
\begin{align}
\sigma_q \equiv \left( \frac{\int dq\, q^4 f_\chi(q)}{\int dq\, q^2 f_\chi(q)} \right)^{1/2} \ .
\label{eq:sigmaq}
\end{align}
Although $T_\chi$ and $\sigma_q$ both depend explicitly on the arbitrary choice of $T_P$, physical results remain unaffected. This is evident from Eq.~\eqref{eq:Wchi}: their product corresponding to the r.m.s. physical momentum is invariant. As we will make explicit in the next paragraph, $W_\chi$ is the key quantity for setting mass bounds, confirming that our results are independent of the choice of $T_P$.

Having established the necessary framework, we now determine DM mass bounds by comparing the warmness $W_\chi$ of a given model to the corresponding $W_{\rm WDM}$ for the WDM scenario. For convenience, all relevant WDM results are summarized in App.~\ref{app:WDM}, including the PSD shape, the relic density constraint that fixes the WDM temperature, and the calculation of $W_{\rm WDM}$ as defined in Eq.~\eqref{eq:Wchi}. The mass bound is obtained as follows. We start from a lower bound $m_{\rm WDM}^{\rm min}$ taken from the list reported in Sec.~\ref{sec:WDM}. This minimum WDM mass corresponds to a maximum warmness $W^{\rm max}_{\rm WDM}$. A quasi-thermal DM candidate must then satisfy $W_\chi \leq W^{\rm max}_{\rm WDM}$, yielding the minimum allowed DM mass
\begin{equation}
\label{eq:warmness_map}
m_\chi^{\rm min} = 22\ {\rm keV} \left( \frac{m_{\rm WDM}^{\rm min}}{6\ {\rm keV}} \right)^{4/3} \left( \frac{\sigma_q}{3.6} \right) \left( \frac{106.75}{g_{\star s}(T_P)} \right)^{1/3} .
\end{equation}
Rescaling a given WDM mass bound $m_{\rm WDM}^{\rm min}$ therefore requires knowledge of two parameters: the dimensionless second moment of the PSD, $\sigma_q$, and the temperature introduced to define the comoving momentum, $T_P$. This might suggest that the resulting mass bound depends on the (arbitrary) choice of $T_P$. However, as already emphasized, physical results cannot depend on this choice. Indeed, the dimensionless second moment $\sigma_q$ depends on the choice of $T_P,$ and this dependence exactly cancels out with that of the $g_{\star s}(T_P)$ factor, rendering the mass bound in Eq.~\eqref{eq:warmness_map} independent of $T_P$. 

\begin{figure}
\centering
\includegraphics[width=.95\linewidth]{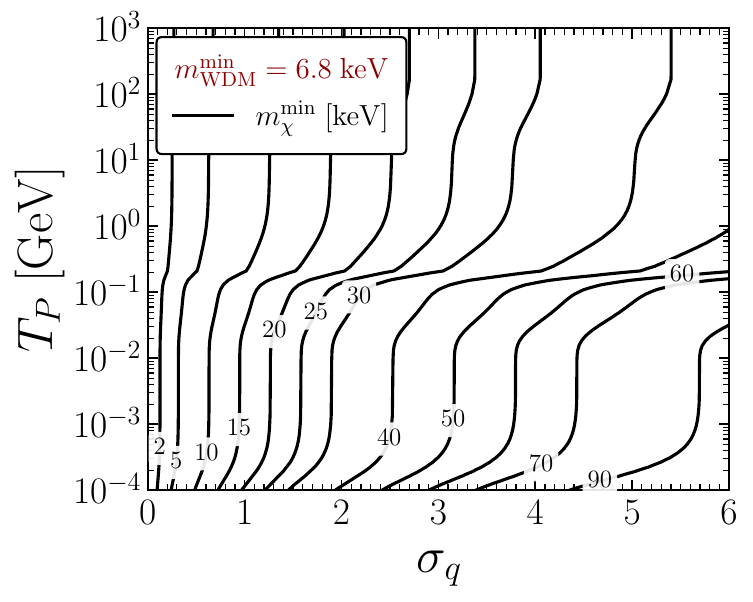}
\caption{Mass bounds on DM with a quasi-thermal PSD in the $(\sigma_q, T_P)$ plane. The dimensionless second moment $\sigma_q$ is defined in Eq.~\eqref{eq:sigmaq}, while the temperature $T_P$ appears in the definition of the comoving momentum, Eq.~\eqref{eq:qcom}. Contours show the minimum allowed DM mass, obtained via the rescaling in Eq.~\eqref{eq:warmness_map} using $m_{\rm WDM}^{\rm min} = 6.8\,{\rm keV}$ as the reference value.}
\label{fig:general_bounds}
\end{figure}

We summarize our results in the general form shown in Fig.~\ref{fig:general_bounds}. Taking $m_{\rm WDM}^{\rm min} = 6.8\ {\rm keV}$ as a reference value, we display isocontours of $m_\chi^{\rm min}$ in the $(\sigma_q, T_P)$ plane. The interpretation of this plane is as follows: given a specific DM model with fixed masses and couplings, one selects a convenient temperature $T_P$ and solves the Boltzmann equation to obtain the PSD. This allows for the computation of $\sigma_q$, identifying a point in the plane. Repeating the procedure with a different $T_P$ leads to a different value of $\sigma_q$, thereby shifting the model’s location in the plane. In summary, each model traces out a line in the $(\sigma_q, T_P)$ plane that follows the shape of the isocontours in Fig.~\ref{fig:general_bounds}, which represent constant values of $W_\chi$.

Associating $T_P$ with the bath temperature at which DM production is most efficient clarifies the behavior of the isocontours in Fig.~\ref{fig:general_bounds}. The weakest mass bounds appear in the upper-left region of the plot. This is expected: moving left along the horizontal axis corresponds to smaller values of $\sigma_q$, indicating colder DM distributions with lower velocity dispersion. The vertical dependence is similarly intuitive: larger values of $T_P$ correspond to earlier production times, resulting in more redshifting and cooling of the DM particles, especially when the thermal bath crosses particle mass thresholds. Notably, the bound becomes significantly more stringent near the GeV scale, where a sharp drop in the number of SM degrees of freedom enhances cooling and leads to a smaller $\sigma_q$.

\section{Freeze-in production}
\label{sec:FI}

After the general discussion about DM candidates with a quasi-thermal PSD, we now turn to the specific production mechanism known as freeze-in. In this scenario, the early universe is assumed to be reheated after inflation with a negligible abundance of DM. Particles in the thermal bath can occasionally scatter or decay into final states that include DM. As long as the relevant interactions are sufficiently feeble, the inverse (destruction) processes remain negligible and DM never reaches thermal equilibrium. The resulting DM population simply redshifts with the expansion of the universe, propagating along geodesics and persisting to the present time.

We follow the notation and conventions of Ref.~\cite{DEramo:2020gpr}, and summarize here only the key results. The general freeze-in production process, involving $n$ bath particles in the initial state and a final state composed of $m$ bath particles and $\ell$ DM particles, reads
\begin{equation}
    \underbrace{\bath_1  + \dots + \bath_n}_{n} \longrightarrow \underbrace{\bath_{n+1} +\dots +\bath_{n+m}}_{m} + \underbrace{\chi + \dots + \chi}_{\ell} \ .
    \label{eq:FIprocess}
\end{equation}
Bath particles $\bath_i$ are described by equilibrium distributions $f^{\rm eq}_{\bath_i}(k_i) = [\exp(E_i/T)\pm 1]^{-1}$, either characterized by Fermi-Dirac ($+$) or Bose-Einstein ($-$) statistics.  The PSD evolution is governed by the Boltzmann equation 
\begin{equation}\label{eq:BE}
    g_\chi\frac{ df_\chi(p, t)}{dt} = g_\chi \frac{\sum_\alpha C_\alpha(t)}{E}\ .
\end{equation}
The sum over $\alpha$ runs over all contributing production processes. For freeze-in, the right-hand side of the Boltzmann equation is a function of the bath temperature $T$ and not a functional of the PSD $f_\chi$. Its explicit form for the process in Eq.~\eqref{eq:FIprocess} is
\begin{align}\nonumber 
  C_{n \rightarrow m \ell}  &\equiv \frac{\ell}{2}\int \prod_{i=1}^{n+m}d {\cal K}_i \prod_{j=2}^{\ell}d\Pi_{j} \, (2\pi)^4 \delta^{(4)}(P_{\rm f} - P_{\rm i}) |{\cal M}_\alpha|^2\\ 
   & \qquad\qquad \times \prod_{i=1}^{n}f_{\bath_i}(k_i)\prod_{j={n+1}}^{n+m}[1\mp f_{\bath_j}(k_j)] \ .
\label{eq:collterm}
\end{align}
Here, $P_{\rm i}$ and $P_{\rm f}$ denote the total four-momenta of the initial and final states, respectively. The measures $d{\cal K}_i$ and $d\Pi_j$ are the Lorentz-invariant phase space elements, given by $g_i d^3 p_i / [(2\pi)^3 2 E_i]$, for bath and DM particles, respectively. The quantity $|{\cal M}_\alpha|^2$ is the squared matrix element for the process, averaged over both initial and final state degrees of freedom. This, along with the number and type of production channels, constitutes the only model-dependent input for freeze-in.

Before solving the Boltzmann equation, it is convenient to change the time and momentum variables. First, we replace the cosmic time $t$ with the thermal bath temperature $T$; the corresponding Jacobian factor is determined from entropy conservation. For the momentum, we adopt the comoving variable introduced in Eq.~\eqref{eq:qcom}, which naturally factors out the effects of cosmic redshift.

The Boltzmann equation is integrated from an initial temperature $T_{\rm UV}$, chosen to be much larger than the reference temperature $T_P$ at which production is most efficient, down to the present temperature $T_0$. This yields the asymptotic comoving distribution
\begin{equation}\label{eq:PSD_solution}
    g_\chi f_\chi = \int_{T_{\rm UV}}^{T_0} \frac{d\log T}{H(T)} \left(1 + \frac{1}{3} \frac{d\log g_{\star s}}{d\log T} \right) g_\chi \frac{\sum_\alpha C_\alpha}{E} \ .
\end{equation}
We impose the initial condition $f_\chi(T_{\rm UV}) = 0$ and focus on scenarios in which the final relic abundance is insensitive to the precise value of $T_{\rm UV}$. It has been shown that this IR dominated production happens always for decays and for scattering mediated by renormalizable interactions~\cite{McDonald:2001vt,Kusenko:2006rh,Ibarra:2008kn,Hall:2009bx}. We safely take the $T_{\rm UV} \rightarrow \infty$ limit in the integral above. Moreover, if we are interested in the asymptotic form of the PSD expressed in terms of the comoving momentum, there is no need to integrate all the way down to the present temperature $T_0$. Upon identifying $T_P$ with the mass of the heaviest particle involved in the production process, it is sufficient to stop the integration at a temperature well below $T_P$, where the Maxwell–Boltzmann exponential suppression of the bath particle density renders further production negligible.

Whether DM production is single (i.e., $\ell = 1$) or multiple (i.e., $\ell > 1$), the functional shape of the PSD resulting from the integral in Eq.~\eqref{eq:PSD_solution} is not substantially affected. The main difference is the multiplicative factor of $\ell$ in the collision term, Eq.~\eqref{eq:collterm}, which is reabsorbed by the normalization imposed by the relic density constraint. Minor corrections may arise from the phase space integrals, as multiple production requires integration over the momenta of the additional DM particles. This paper focuses on the regime where DM is much lighter than the heaviest bath particle involved in production. In this regime, and absent nearly degenerate bath particles, integrating over additional DM momenta is equivalent to integrating over momenta of light bath particles. While the detailed functional form may differ, in the IR-dominated regime the matrix element can be taken as constant to excellent approximation. Pauli-blocking or Bose–Einstein effects for final-state bath particles introduce corrections to the second moment $\sigma_q$ at the percent level~\cite{DEramo:2020gpr}. For these reasons, we concentrate here on single production, while multiple production is treated in the next section in the context of explicit microscopic models.

\begin{figure*}
\centering
\includegraphics[width=1\linewidth]{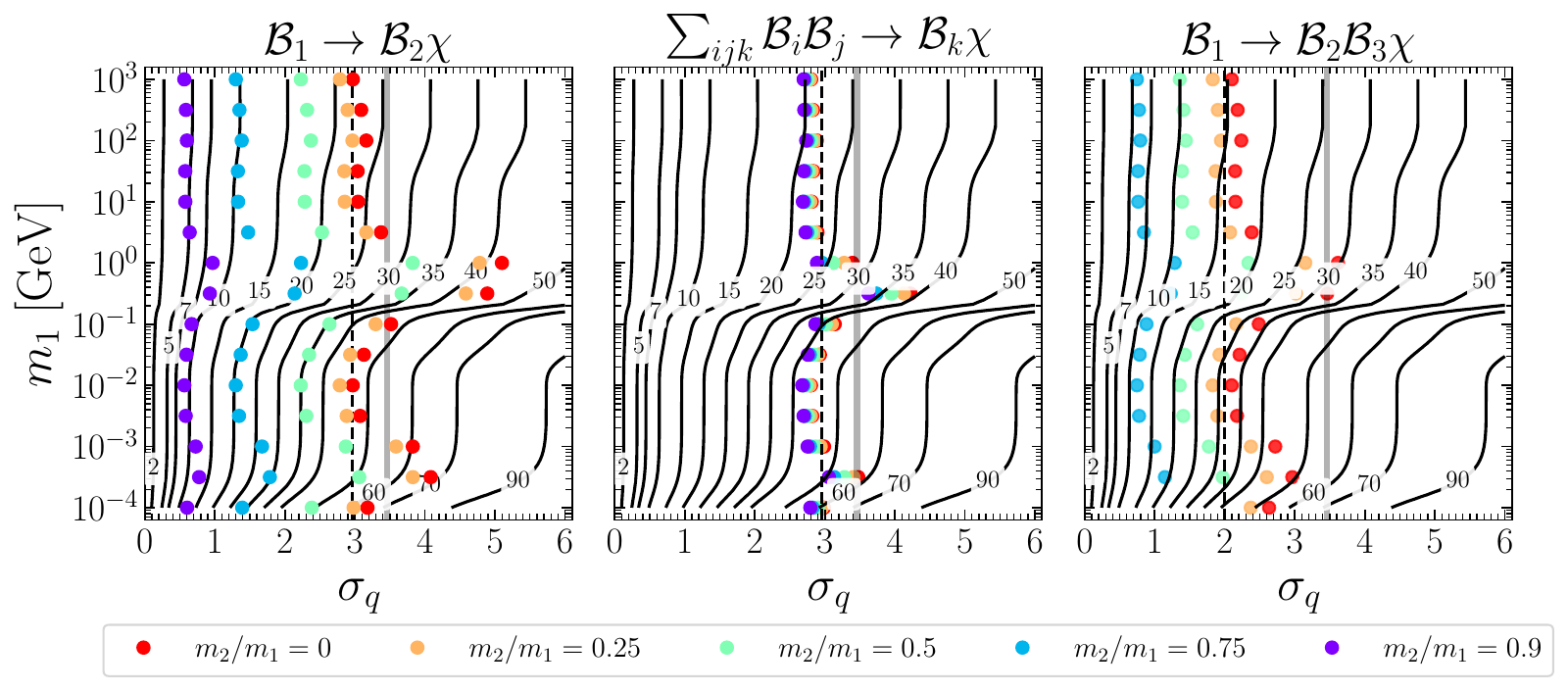}
\caption{Lower mass bounds on freeze-in produced DM from: two-body decays (left), binary scatterings (center), and three-body decays (right). The $(\sigma_q, m_1)$ plane is the same as in Fig.~\ref{fig:general_bounds}, with $T_P$ replaced by the mass of the heaviest particle involved in production, $m_1$. Solid black contours reproduce those of Fig.~\ref{fig:general_bounds}, corresponding to the reference WDM bound $m^{\rm min}_{\rm WDM} = 6.8$~keV. For each channel, we explore different mass spectra, as indicated in the legend, by varying $m_2/m_1 = \{0, 0.25, 0.5, 0.75, 0.9\}$. Whenever a third bath particle is involved, we always set $m_3 = 0$. Each choice for the mass spectrum identifies a point in the plane; the black contour intersecting the point gives the associated mass bound. Vertical dashed lines show analytically computed values of $\sigma_q$ (see App.~\ref{app:Cs}), and the thick gray line marks the WDM benchmark $\sigma_q^{\rm WDM} \simeq 3.6$.
} 
\label{fig:phase-space-suppression}
\end{figure*}

In what follows, we analyze three distinct freeze-in production channels: two-body decays, binary scatterings, and three-body decays. The analysis in this section is model-independent and depends on a single parameter: the overall freeze-in mass scale, set by the heaviest bath particle involved in the production process. Without loss of generality, we denote this particle as $\bath_1$, with mass $m_1$, thereby defining the production temperature as $T_P = m_1$. The interaction strength is parameterized through a constant matrix element, whose value is fixed by the relic density constraint. All particle statistics are taken to follow Maxwell–Boltzmann distributions. For each production channel, we solve the Boltzmann equation and extract the corresponding value of $\sigma_q$.

Our results are presented in Fig.~\ref{fig:phase-space-suppression}. For each case, the masses of the other bath particles are chosen such that $m_2/m_1$ is set by the legend, and $m_3 = 0$. The background (black) contours in all three panels reproduce those of Fig.~\ref{fig:general_bounds}, which show DM mass bounds in the $(\sigma_q, T_P)$ plane with the only difference that here the vertical axis is relabeled as $m_1$. A specific choice for the mass spectrum corresponds to a colored point in this plane. In the evaluation of the collision terms, the DM mass is chosen to satisfy $m_\chi \ll m_1$ and is taken to lie slightly above the corresponding bound.

\begin{itemize}
\item \textbf{Two-body decays.} The case of two body decays $\bath_1 \to \bath_2 \chi$ is the only one in which the constant matrix element approximation holds exactly. We observe that, except in the region where $m_1$ lies between $0.1$ and $1$~GeV and near the MeV scale, the bounds on $m_\chi$ are less stringent compared to the thermal scenario, indicated by the vertical thick gray line. Red points correspond to $m_2 = 0$, while purple points indicate $m_2 = 0.9\, m_1$, which reduces the energy available for DM particles. This reduction, referred to as \textit{phase space suppression}, weakens the warmness constraints as expected.
\item \textbf{Binary scattering.} In this case, we consider DM production through three scattering processes ($\bath_1 \bath_2 \to \bath_3 \chi$ and its permutations). All matrix elements are approximated as constants. We present results in the $(\sigma_q, m_1)$ plane for different choices of the mass spectrum once the largest scale $m_1$ is fixed. The color coding corresponds to different values of the mass ratio $m_2/m_1$. The results correctly reflect that phase space suppression cannot occur in this case, since we include all permutations of the initial and final state bath particles. The bounds on $m_\chi$ are again weaker compared to the thermal PSD case. Notably, we find that the mass bounds depend only mildly on $m_2$ and $m_3$, with the dominant factor being $m_1$, which sets the typical production energy. For $m_1$ greater than a GeV, we find $m_{\chi}^{\rm min} \simeq 17$~keV, while for $m_1$ below $100$~MeV, we obtain $m_{\chi}^{\rm min} \simeq 30$~keV. 
\item \textbf{Three-body decays.} The case of three body decays $\bath_1 \to \bath_2 \bath_3 \chi$ shares features with both scenarios discussed above. We again approximate the matrix elements as constants. Since this is a decay, the total energy in the final state is constrained by the mass of the decaying particle. As a result, warmness bounds can be relaxed by phase space suppression, achieved by choosing $m_2$ sufficiently close to $m_1$. However, even when $m_2$ is large, the resulting bounds are generally weaker than those obtained for two-body decays. Specifically, for $m_1$ greater than a GeV, we find $m_{\chi}^{\rm min} \simeq 13$~keV, while for $m_1$ below $100$~MeV, we obtain $m_{\chi}^{\rm min} \simeq 27$~keV.
\end{itemize}

\begin{figure*}
\centering
\includegraphics[width=1\linewidth]{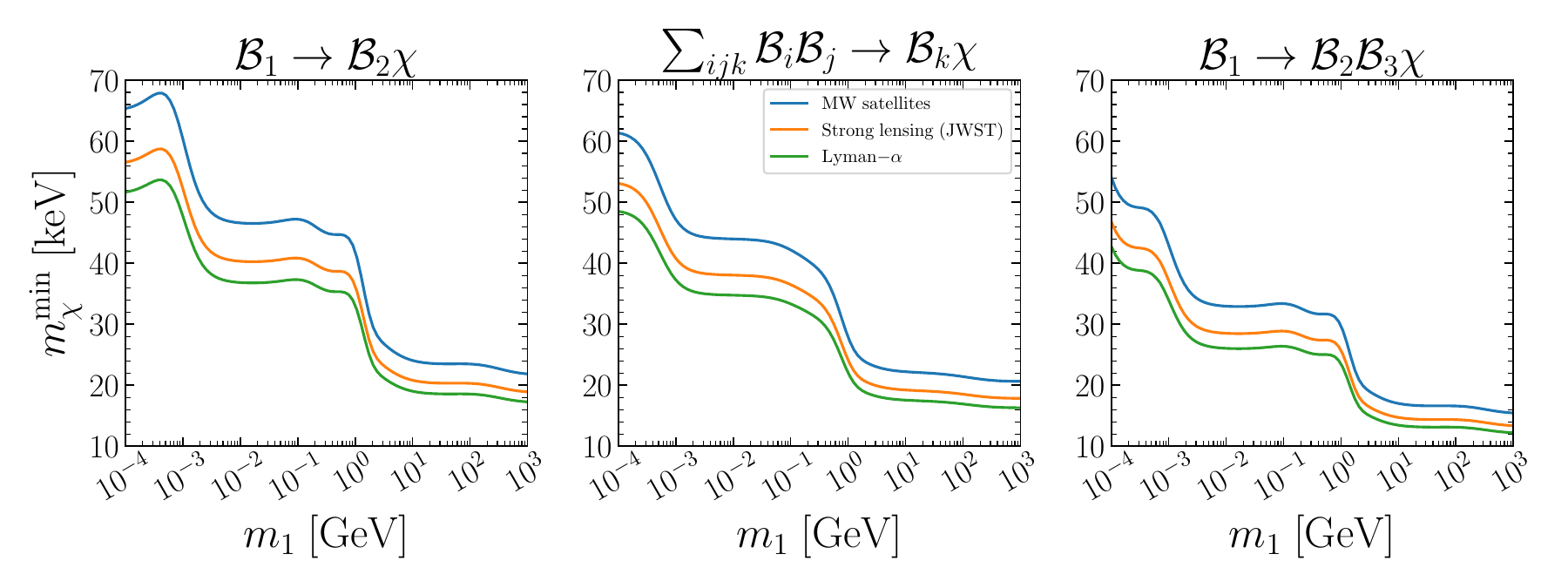}
\caption{Lower bounds on the DM mass $m_{\chi}^{\rm min}$ for the three production mechanisms considered in this work: two body decays (left), binary scatterings (center), and three body decays (right). The three lines in each panel correspond to the reference WDM mass bounds shown in Fig.~\ref{fig:wdm-consraints}, using the same color scheme. Matrix elements are taken to be constant and adjusted to reproduce the observed DM relic abundance. The variable $m_1$ denotes the mass of the heaviest particle participating in the process, while $\bath_2$ (and $\bath_3$ when present) are assumed to be massless. The DM mass bound is shown as a function of $m_1$.} 
\label{fig:observables_topologies}
\end{figure*}

Finally, we derive lower bounds on the DM mass $m_\chi$ using Eq.~\eqref{eq:warmness_map}, taking into account different values for the WDM mass bound $m^{\rm min}_{\rm WDM}$. The resulting constraints are displayed in Fig.~\ref{fig:observables_topologies} for the three production mechanisms discussed above, with $m_1$ varied and all other particles taken to be massless (except the DM). Among these scenarios, two-body decays yield the strongest bounds, followed by binary scatterings and then three-body decays.

\section{Minimal two-body decays}
\label{sec:UV1}

After the model independent analysis of freeze-in presented in the previous section, we now derive mass bounds for explicit examples of microscopic theories. In this section, we focus on two \textit{minimal} scenarios, each based on a renormalizable extension of the SM, that offer a viable DM candidate produced via two-body decays. It is important to recall that the results of the previous section were obtained under two simplifying assumptions: constant matrix elements and Maxwell-Boltzmann statistics for all particles involved. The first assumption is well justified in the case of two-body decays, as the final state is monochromatic. Once a specific microscopic model is chosen, however, we can go beyond these approximations by solving the Boltzmann equation with the exact collision operators and including all quantum statistical effects. As shown in Ref.~\cite{DEramo:2020gpr}, these quantum corrections induce only a small change in the parameter $\sigma_q$ defined in Eq.~\eqref{eq:sigmaq}, typically at the percent level. Consequently, the resulting shift in the mass bound derived from Eq.~\eqref{eq:warmness_map} is negligible. This confirms that the model independent results obtained previously remain reliable even in the context of complete microscopic models.

\subsection{Scalar DM via the Higgs portal}

The first case we consider is the most minimal extension of the SM in terms of new degrees of freedom. We introduce one real scalar field $\phi$, neutral under all SM gauge symmetries. To ensure the stability of $\phi$, we impose a discrete $\mathbb{Z}_2$ symmetry under which all SM fields are even and $\phi$ is odd. The most general set of renormalizable interactions involving $\phi$ that can be added to the SM Lagrangian is given by
\begin{equation}
\mathcal{L}_\phi = \frac{1}{2} (\partial_\mu \phi)^2 - \frac{1}{2} m_\phi^2 \phi^2  - \frac{\lambda_{\rm mix}}{2} H^\dag H \phi^2 - \frac{\lambda_\phi}{4!} \phi^4 \ .
\end{equation}
The first two terms describe a canonically normalized real scalar with mass $m_\phi$. To preserve the $\mathbb{Z}_2$ symmetry and guarantee the stability of $\phi$, we must prevent it from acquiring a vacuum expectation value (vev). In the regime $\lambda_{\rm mix} \ll 1$, which is required for freeze-in production, the scalar potential can be analyzed by decoupling the dynamics of the Higgs doublet. In this limit, the condition for a vanishing vacuum expectation value is simply $\lambda_\phi > 0$ and $m_\phi^2 > 0$.

DM production in the early universe is controlled by the interaction between the field $\phi$ and the SM Higgs boson $h$, identified as $H = \left(0 \;\; (v_H + h)/\sqrt{2}\right)^T$, where the Higgs vev is determined from the Fermi constant and evaluates to $v_H \simeq 246\, \mathrm{GeV}$. The relevant production channels for DM are mediated by the cubic interaction term $h \phi^2$, which emerges from expanding the scalar potential around the Higgs vev. This setup has been shown to reproduce the observed DM abundance via both freeze-out~\cite{Burgess:2000yq,McDonald:1993ex} and freeze-in~\cite{McDonald:2001vt}. The freeze-out regime is tightly constrained by direct detection experiments.

For DM masses below half the Higgs mass, the decay $h \rightarrow \phi \phi$ is kinematically allowed, with decay width
\begin{equation}
\Gamma_{h \rightarrow \phi \phi} = \frac{\lambda_{\rm mix}^2 v_H^2}{32 \pi m_h} 
\sqrt{1 - \frac{4 m_\phi^2}{m_h^2}} \, .
\end{equation}
Since we focus on freeze-in production of DM in the tens of keV range, this decay channel is always open. Moreover, it has been shown that the contribution from scatterings is subdominant and can be safely neglected when Higgs decays are kinematically accessible~\cite{Bernal:2018kcw,DEramo:2024lsk}. This setup thus provides a minimal and concrete scenario in which DM is produced via freeze-in through the two-body decay of a SM particle with known mass.

The squared matrix element for a two-body decay is constant and directly proportional to the decay width. We fix its value (i.e., the coupling $\lambda_{\rm mix}$) to reproduce the observed DM relic density, leaving the DM mass as the only free parameter subject to constraints. The lower bound on $m_\phi$ can be inferred from the left panel of Fig.~\ref{fig:observables_topologies}, selecting $m_1 = m_h \simeq 125 \, {\rm GeV}$, with the caveat that those results were obtained under the assumption of Maxwell-Boltzmann statistics and constant matrix elements. However, quantum statistical corrections affect the mass bound only at the percent level, making the Maxwell-Boltzmann approximation sufficiently accurate. The resulting bounds on the DM mass are
\begin{equation}
m_\phi > \begin{cases}
    23 \, {\rm keV} & \text{MW satellites} \\
    20 \, {\rm keV} & \text{JWST lensing} \\
    18 \, {\rm keV} & \text{Lyman}\,\alpha \\
\end{cases} \ .
\end{equation}

\subsection{Sterile Neutrinos}

The second model we study is also guided by the criterion of minimality, but now we introduce a new fermionic degree of freedom in the form of a Weyl field $N$. As in the previous case, the new field is assumed to be neutral under all SM gauge symmetries. This choice significantly complicates the construction of a viable Lagrangian if we continue to restrict ourselves to renormalizable theories. The most general set of renormalizable terms that can be added to the SM Lagrangian reads
\begin{equation}
\mathcal{L}_{N} = N^\dag i \overline{\sigma}^\mu \partial_\mu N - \left[\frac{m_N}{2} N N + \lambda_\nu L H N + {\rm h.c.} \right] \ . 
\label{eq:LN}
\end{equation}
In addition to the canonical kinetic term, there are two contributions: a Majorana mass term, which is allowed since $N$ is a complete SM singlet, and a Yukawa interaction involving the SM gauge-invariant combination $L H$, where $L$ denotes the lepton doublet. We omit flavor indices in the Lagrangian above. In principle, one could consider scenarios in which $N$ couples to all SM lepton generations, promoting the Yukawa coupling to a vector, or even introduce multiple fermion singlets, resulting in a full Yukawa matrix. However, the conclusions of this subsection are not sensitive to these model-building choices.

The scenario illustrated by the above Lagrangian is precisely what is required to generate masses for the SM \textit{active neutrinos} through the addition of fermion singlets. Moreover, if the \textit{sterile neutrino} field $N$ has a mass in the tens of keV range, it remains stable on cosmological timescales and can serve as a viable DM candidate. An unavoidable production mechanism for sterile neutrinos in the early universe is the Dodelson-Widrow process~\cite{Dodelson:1993je}, in which oscillations of active neutrino fields $\nu$ populate the sterile sector. However, the scenario in which sterile neutrino DM is produced solely via this mechanism is excluded by X-ray observations, which constrain the radiative decay channel $N \rightarrow \gamma \nu$ by searching for monochromatic photon lines~\cite{Dessert:2023fen,Abazajian:2017tcc}. A possible way to evade these bounds is to invoke an alternative DM genesis via resonant production through the Shi-Fuller mechanism~\cite{Shi:1998km} or non-standard neutrino self-interactions~\cite{DeGouvea:2019wpf,Kelly:2020pcy}.

In this work, we consider another production channel. The Yukawa interaction in Eq.~\eqref{eq:LN} can be made invariant under lepton number if the field $N$ is appropriately charged. However, the Majorana mass term violates lepton number by two units. This low-energy theory can arise from a lepton-number-conserving ultraviolet completion by introducing a new complex scalar $\Phi$, which is also a singlet under the SM gauge group. The Yukawa operator $\Phi N N$ conserves lepton number provided $\Phi$ carries the appropriate charge. If $\Phi$ acquires a nonvanishing vev, it generates the Majorana mass term for $N$ at low energies. Expanding the scalar field around its vev as $\Phi = (v_\Phi + \rho)/\sqrt{2}$, we identify the radial mode $\rho$, which can thermalize with the SM bath through a quartic interaction with the Higgs doublet $H$. The decay of this thermalized $\rho$ field into sterile neutrinos then acts as a freeze-in production mechanism. We focus on the regime where this is the dominant channel for sterile neutrino DM production~\cite{Konig:2016dzg}. Unlike the minimal Higgs portal scenario discussed earlier, here the mass of the decaying particle $m_\rho$ remains a free parameter. We derive the following lower bounds on the DM mass
\begin{equation}
m_N > 
\left\{
\begin{array}{ll}
\begin{cases}
47\,{\rm keV} & \text{MW Satellites} \\
40\,{\rm keV} & \text{JWST} \\
37\,{\rm keV} & \text{Lyman}\mathrm{-}\alpha 
\end{cases}
& \quad m_\rho = 100\,{\rm MeV} \\[3em]
\begin{cases}
24\,{\rm keV} & \text{MW Satellites} \\
21\,{\rm keV} & \text{JWST} \\
19\,{\rm keV} & \text{Lyman}\mathrm{-}\alpha 
\end{cases}
& \quad m_\rho = 10\,{\rm GeV} \\[3em]
\begin{cases}
22\,{\rm keV} & \text{MW Satellites} \\
19\,{\rm keV} & \text{JWST} \\
17\,{\rm keV} & \text{Lyman}\mathrm{-}\alpha 
\end{cases}
& \quad m_\rho = 1\,{\rm TeV}
\end{array}
\right. \ .
\end{equation}
As in the previous scenario, we verified that including the Bose-Einstein statistical factors for the decaying scalar alters the mass bounds only at the percent level.

\begin{table*}
\centering
\begin{tabular}{c||c|c|c||c|c|c|c|}
\multicolumn{1}{c||}{} & \multicolumn{3}{c||}{$m_a^{\rm min}$(keV)} & \multicolumn{4}{c|}{Relic density requirement and X-ray constraints} \\
\hline 
$\psi$ & Ly$\alpha$ & JWST & Satellite & $\Lambda_\psi^{\rm relic}$(GeV) & $-\Delta g_{a\gamma\gamma}$ (GeV$^{-1}$) & $\tau_{a\rightarrow\gamma\gamma}$(sec) & Allowed? \\
\hline\hline
electron & $39$ & $43$ & $49$ & $5.7\times  10^7$ & $3.1 \times 10^{-14}$ & $1.2\times 10^{18}$ & \color{nicered}{\ding{55}} \\
muon & $33$ & $37$ & $43$ & $3.6\times 10^8$ & $5.6\times10^{-20}$ & $1.1 \times  10^{30}$ & \color{nicegreen}{\ding{51}} \\
tau & $19$ & $21$ & $24$ & $4.9\times 10^8$ & $7.3\times 10^{-23}$ & $1.8\times 10^{36}$ & \color{nicegreen}{\ding{51}} \\
\hline
charm & $19$ & $20$ & $23$ & $4.5\times 10^9$ & $1.8\times 10^{-23}$ & $3.1\times 10^{37}$ & \color{nicegreen}{\ding{51}} \\
bottom & $19$ & $21$ & $24$ & $7.3 \times 10^9$ & $3.2\times 10^{-25}$ & $8.3 \times 10^{40}$ & \color{nicegreen}{\ding{51}} \\
top & $18$ & $19$ & $22$ & $2.8\times 10^{10}$ & $1.5\times 10^{-28}$ & $5.5\times 10^{47}$ & \color{nicegreen}{\ding{51}} \\
\hline
\end{tabular}
\caption{Summary of the results for the ALP DM scenario in Eq.~\eqref{eq:LaFC} with flavor-conserving couplings to SM fermions. Freeze-in production takes place via binary scatterings. We consider both couplings to charged leptons (first three rows) and to heavy quarks (last three rows). For each choice of the SM fermion $\psi$, we find the mass bounds $m_a^{\rm min}$ from three different observables and report their values. For ALP masses in this ballpark range, the DM relic density constraint fixes the value of the axion coupling to $\Lambda^{\rm relic}_\psi$. We also determine the radiatively induced coupling $g^{\rm 1-loop}_a$ to photons and the resulting ALP lifetime $\tau_{a\rightarrow\gamma\gamma}$. Finally, in the last column we state whether this scenario is allowed by X-ray constraints.}\label{tab:FC}
\end{table*}

\section{ALP coupled to SM fermions}
\label{sec:UV2}

We now consider scenarios where DM production occurs via both two-body decays and binary scatterings. The DM candidate is a CP-odd scalar $a$, originating from the spontaneous breaking of a global symmetry. A prominent example is the QCD axion associated with Peccei-Quinn symmetry breaking~\cite{Peccei:1977np,Peccei:1977ur,Wilczek:1977pj,Weinberg:1977ma}. Adopting a low-energy effective field theory, we impose a shift symmetry consistent with the Nambu-Goldstone nature of $a$, and focus on couplings of this ALP to SM fermions only
\begin{equation}
\mathcal{L}_a = \frac{1}{2} (\partial_\mu a)^2 - \frac{1}{2} m_a^2 a^2 + \mathcal{L}_a^{\rm (FC)} + \mathcal{L}_a^{\rm (FV)} \ .
\label{eq:LaFC}
\end{equation}
Here, $\mathcal{L}_a^{\rm (FC)}$ and $\mathcal{L}_a^{\rm (FV)}$ contain flavor-conserving and violating interactions, respectively. The mass term softly breaks the shift symmetry and is a free parameter.

\subsection{Flavor-conserving couplings}

We consider the flavor-conserving interactions defined in the basis where the SM Yukawa matrices are diagonal
\begin{equation}
\mathcal{L}_a^{\rm (FC)} = - \frac{\partial_\mu a}{2 \Lambda}\sum_\psi \mathcal{C}_\psi  \overline \psi \gamma^\mu \gamma^5 \psi \ .
\label{eq:LaFV}
\end{equation}
It is convenient to define the combination $\Lambda_\psi \equiv \Lambda/ \mathcal{C}_\psi$. 

For ALP coupled to leptons, $\psi = \ell$, the production channels are lepton/antilepton annihilations $\ell^+ \ell^- \rightarrow \gamma a$ and Compton-like scattering $\ell^{\pm} \gamma \rightarrow \ell^{\pm} a$. The squared matrix elements expressed in terms of the Mandelstam variables result in~\cite{DEramo:2018vss,Badziak:2024qjg,DEramo:2024jhn}
\begin{subequations}
\begin{align}
|{\cal M}_{\ell^+ \ell^- \to a \gamma }|^2 = & \, \frac{e^2}{2 \Lambda_\ell^2} 
\frac{m_\ell^2 s^2}{(m_\ell^2 - t)(s+t - m_\ell^2)} \ ,\\
|{\cal M}_{\ell^\pm \gamma \to \ell^\pm a }|^2 = & \, \frac{e^2}{2 \Lambda_\ell^2} \frac{m_\ell^2 t^2}{(s-m_\ell^2)(s+t - m_\ell^2)} \ ,
\end{align}
\end{subequations}
where $e$ is the electron charge. 

The analysis for ALPs coupled to quarks proceeds analogously, with the key difference that we restrict to the three heaviest quarks to avoid issues due to confinement. The relevant squared matrix elements are
\begin{subequations}
\begin{align}
|{\cal M}_{q \bar{q} \to g a}|^2 &= \frac{g_s^2}{36 \Lambda_q^2} \frac{m_q^2 s^2}{(m_q^2 - t)(s + t - m_q^2)} \ , \\
|{\cal M}_{q g \to q a}|^2 &= \frac{g_s^2}{36 \Lambda_q^2} \frac{m_q^2 t^2}{(s - m_q^2)(s + t - m_q^2)} \ ,
\end{align}
\end{subequations}
where $g_s$ denotes the strong coupling constant. For the top quark, whose mass lies well above the QCD scale, freeze-in calculations are fully under control. For the bottom and charm quarks, we conservatively stop the Boltzmann evolution at $T_{\rm stop} = 1\,{\rm GeV}$ and neglect production at lower temperatures. Similar issues have been discussed in the context of axion dark radiation~\cite{Arias-Aragon:2020shv,DEramo:2024jhn}, where thermal axions lead to potentially observable contributions to $\Delta N_{\rm eff}$. Here, by contrast, we are interested only in the \textit{shape} of the resulting PSD, which depends on the temperature evolution of each individual momentum bin in the collision term. Ref.~\cite{DEramo:2023nzt} showed how this dependence is mild, supporting the reliability of our results.

ALP-fermion interactions induce one-loop couplings to photons, rendering the ALP unstable and enabling observational tests through X-ray line searches. The effective interaction generated at one loop is given by~\cite{Bauer:2017ris,Bauer:2020jbp}
\begin{subequations}
\begin{align}
\Delta\mathcal{L}^{\rm (1-loop)}_a &= - \frac{\Delta g_{a\gamma\gamma}}{4} \, a \, F^{\mu\nu} \widetilde{F}_{\mu\nu} \ , \\
\Delta g_{a\gamma\gamma} &\simeq - \frac{\alpha_{\rm em} N^c_\psi Q_\psi^2}{12 \pi \Lambda_\psi} \left( \frac{m_a}{m_\psi} \right)^2 \ .
\end{align}
\end{subequations}
This leads to a decay width $\Gamma_{a \rightarrow \gamma \gamma} = \Delta g_{a\gamma\gamma}^2 m_a^3 / (64 \pi)$. We summarize the resulting constraints for ALP DM produced via flavor-conserving scatterings with SM fermions in Tab.~\ref{tab:FC}. Notably, the case of ALPs coupled to electrons is excluded by X-ray observations \cite{Cadamuro:2011fd,Perez:2016tcq,Ng:2019gch,Roach:2022lgo}. 

\begin{table}
\centering
\begin{tabular}{c||c|c|c||c|}
\multicolumn{1}{c||}{} & \multicolumn{3}{c||}{$m_a^{\rm min}$(keV)} & \\
\hline 
$\psi$ & Ly$\alpha$ & JWST & Satellite & $\Lambda_{\psi\psi^\prime}^{\rm relic}$(GeV)  \\
\hline\hline
tau & 25 & 27 & 31 & $2.9\times 10^9$ \\
muon & 37 & 40 & 47 & $2.9\times 10^9$ \\
\hline
top & $18$ & $20$ & $22$ & $3.4\times 10^{10}$ \\
\hline
\end{tabular}
\caption{Results for the ALP DM scenario described by Eq.~\eqref{eq:LaFV}, with flavor-violating couplings to SM fermions. Freeze-in production occurs via two-body decays $\psi \rightarrow \psi^\prime a$. We consider couplings to charged leptons and the top quark. Decays of bottom and charm quarks are omitted for reasons discussed in the main text. For each SM fermion $\psi$, we report the lower bound on the ALP mass $m_a^{\rm min}$ from three different observables, along with the value of the coupling scale $\Lambda^{\rm relic}_{\psi\psi^\prime}$ defined in Eq.~\eqref{eq:Lambdapsipsiprime} required to reproduce the relic abundance.}
\label{tab:FV}
\end{table}

\subsection{Flavor-violating couplings}

We now consider the case in which the ALP couplings are flavor-violating with interactions
\begin{equation}
\mathcal{L}_a^{\rm (FV)} = - \frac{\partial_\mu a}{2 \Lambda}  \sum_{\psi \neq \psi^\prime} \overline{\psi^\prime}\gamma^\mu \left({\cal V}_{\psi \psi^\prime}  +{\cal A}_{\psi \psi^\prime}  \gamma^5\right)\psi + {\rm h.c.} \ ,
\label{eq:LaFV}
\end{equation}
where ${\cal V}_{\psi \psi^\prime}$ and ${\cal A}_{\psi \psi^\prime}$ encode the vector and axial-vector couplings, respectively. These interactions have been considered to exploit the consequences of flavor-violating couplings not only for ALP DM~\cite{Panci:2022wlc,Aghaie:2024jkj} but also for ALP dark radiation~\cite{DEramo:2021usm}. We define the effective scale 
\begin{equation}
    \Lambda_{\psi \psi^\prime} \equiv \frac{\Lambda}{\sqrt{\left|{\cal V}_{\psi \psi^\prime}\right|^2 + \left|{\cal A}_{\psi \psi^\prime}\right|^2}} \ .
\label{eq:Lambdapsipsiprime}
\end{equation}
The squared matrix element for $\psi \to \psi^\prime a$ reads~\cite{DEramo:2018vss,Arias-Aragon:2020shv}
\begin{align}
|{\cal M}_{\psi \to \psi^\prime a}|^2 = \frac{m_\psi^4}{8 N_c^\psi \Lambda_{\psi\psi^\prime}^2} \left( 1 - \frac{m_{\psi^\prime}^2}{m_{\psi}^2} \right)^2 \ ,
\end{align}
with $N_c^\psi$ the number of colors of the fermion $\psi$. 

We present the results for ALP production via flavor-violating decays of SM fermions in Tab.~\ref{tab:FV}. The cases involving charged leptons pose no complications from strong interactions and yield reliable mass bounds, fully consistent with our model-independent analysis. Quark decays, on the other hand, are more delicate due to confinement effects. The result for the top quark, owing to its large mass, remains robust and in agreement with the general analysis. The situation differs significantly for bottom and charm quarks. As in the case of freeze-in production via scatterings with flavor-conserving couplings, we adopt a conservative approach by halting the Boltzmann integration at $T_{\rm stop} = 1\,{\rm GeV}$ to avoid uncertainties near the QCD phase transition. This leads to surprisingly weaker mass bounds, especially for the charm quark, compared to the model-independent expectations. However, this discrepancy is not unexpected. As shown in Ref.~\cite{DEramo:2023nzt}, the collision term’s temperature dependence at fixed momentum is quite pronounced for decays, unlike in the scattering case. In particular, high-momentum modes (relevant at lower temperatures) are artificially removed by the choice $T_{\rm stop}=$ 1 GeV, resulting in a colder and narrower PSD, which in turn relaxes the mass bounds. These results are not reliable, as they are sensitive to the detailed \textit{shape} of the PSD, which cannot be accurately determined when the integration is artificially truncated. For this reason, we choose to omit the bottom and charm quark cases from Tab.~\ref{tab:FV}.

\section{Dark Photon Portal}
\label{sec:UV3}

The last theory we consider features a SM-singlet Dirac fermion $\chi$ as the DM candidate, which interacts with the visible sector through a new spin-one boson $A'_\mu$. The interaction arises via kinetic mixing between the Abelian field strengths of the dark photon (DP) mediator, $F'_{\mu\nu}$, and the SM photon, $F_{\mu\nu}$~\cite{Dvorkin:2019zdi, Chang:2019xva, Dvorkin:2020xga, Bhattiprolu:2024dmh}. The Lagrangian relevant for the phenomenology of this framework is
\begin{equation}
\begin{split}
\mathcal{L}_{\rm DP} = & \,
- \frac{1}{4} F'_{\mu\nu} F'^{\mu\nu} + \frac{\kappa}{2} F'_{\mu\nu} F^{\mu\nu} + \frac{1}{2} m_{A'}^2 A'_\mu A'^\mu \\
& + e J^\mu_{\rm EM} A_\mu + \overline{\chi} \gamma^\mu (i \partial_\mu + e_D A'_\mu) \chi \,.
\end{split}
\end{equation}
Here, $A_\mu$ denotes the SM photon, which couples to the electromagnetic current $J^\mu_{\rm EM}$ composed solely of SM fields, since $\chi$ is a gauge singlet. The Abelian gauge couplings of the visible and dark sectors are denoted by $e$ and $e_D$, respectively.

After transforming to a canonical basis, the DM field acquires a millicharge $Q_\chi = \kappa e_D / e$. This induces an effective interaction that mediates the DM pair-production process $e^+ e^- \to \overline{\chi} \chi$, where we focus on electron–positron annihilations, as heavier fermions give negligible contributions. The squared matrix element, averaged over initial and final spins, is given by
\begin{align}
|{\cal M}_{e^{+}e^{-}\to \overline \chi \chi}|^2 = & \, \frac{Q_\chi^2 e^4}{2s^2} \Big( s^2 + 2st - t(m_e^2 + m_\chi^2) \nonumber \\
& \qquad + 2t^2 + 2(m_e^2 + m_\chi^2)^2 \Big) \,.
\end{align}
A competing process in DM production is plasmon decay. In the thermal plasma, the photon acquires an effective mass from thermal corrections, allowing the decay $\gamma^* \to \overline{\chi} \chi$ to occur. The calculation of the corresponding production rate requires special care, and we adopt the results of Ref.~\cite{Dvorkin:2019zdi}.

\begin{figure}
    \centering
\includegraphics[width=\linewidth]{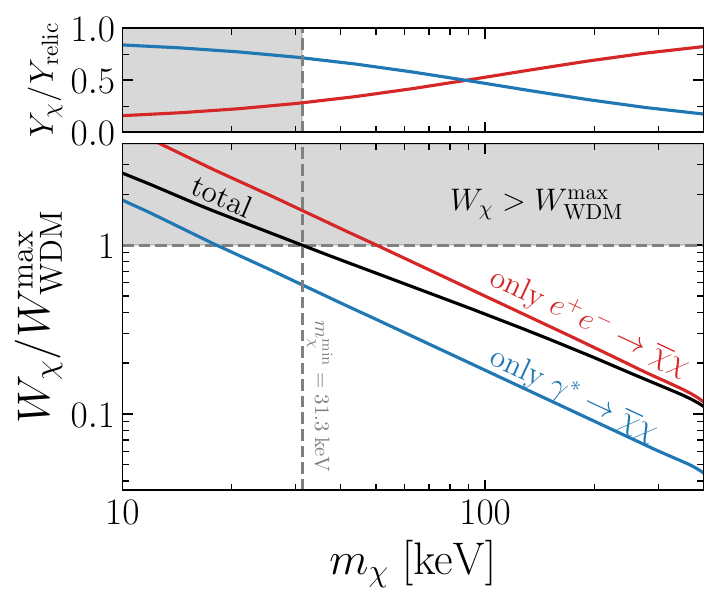}
    \caption{Lower mass bounds for the dark photon portal. The bottom panel shows the ratio $W_\chi / W_{\rm WDM}$ as a function of the DM mass, highlighting the individual contributions from plasmon decay and $e^+e^-$ annihilation. The upper panel displays the corresponding contributions to the relic density. While the relic abundance is predominantly set by plasmon decays, the lower mass bound $m_\chi^{\rm min} = 31.3\,\mathrm{keV}$ (obtained via the condition $W_\chi / W_{\rm WDM} < 1$ and identified by the vertical gray dashed line) lies between the regions where the two production mechanisms dominate. The reference WDM constraint used is $m_{\rm WDM}^{\rm min} = 6.8\,\mathrm{keV}$.}
\label{fig:dark_photon}
\end{figure}

Once both production channels are taken into account, the coupling $Q_\chi$ can be fixed by requiring that $\chi$ reproduces the observed relic abundance. Interestingly, this model allows for DM production via freeze-in through two distinct and competing processes with very different characteristics. While the number density is always dominated by one of the two, making the other a subleading correction, the impact on the phase-space distribution is more subtle. Accounting for both $e^+e^-$ annihilations and plasmon decays is essential. The PSD resulting from plasmon decays is very narrow and peaked at low comoving momenta, whereas the distribution produced by $e^+e^-$ annihilations is much broader and characterized by a higher average momentum. As a result, even a small contribution from annihilations—particularly relevant in the low-$m_\chi$ regime—can significantly ``warm up'' the PSD, affecting the bound on $m_\chi$ at the level of 50\%. Another noteworthy feature of this model is that the production scale depends explicitly on the DM mass. Since plasmon decays occur when the plasma frequency satisfies $\omega_p(T) \simeq 2 m_\chi$, the production temperature is approximately $T \approx 20\, m_\chi$. This behavior is somewhat unconventional for DM produced via IR freeze-in.

Fig.~\ref{fig:dark_photon} illustrates how to identify the DM mass bound. The upper panel displays the relative contributions to the relic density from DM produced via electron–positron annihilation (red) and plasmon decay (blue), as a function of the DM mass. The lower panel shows the ratio between the r.m.s. velocity of the $\chi$ particles and that of the reference WDM model. We show the individual contributions as well as the result obtained from the full PSD. The mass bound is obtained by identifying the intersection point where $W_\chi / W_{\rm WDM} = 1$, and we find
\begin{equation}
m_\chi > \begin{cases}
    31 \, {\rm keV} & \text{MW satellites} \\
    26 \, {\rm keV} & \text{JWST lensing} \\
    24\, {\rm keV} & \text{Lyman}\,\alpha \\
\end{cases} \ .
\end{equation}

We conclude this section with a comparison from the analysis of Ref.~\cite{Dvorkin:2020xga} on the same model. Our bounds are roughly a factor of two more stringent, and we identify this discrepancy as due to different implementations of the criterion used to determine the minimum value of $m_\chi$ consistent with WDM clustering bounds. In Ref.~\cite{Dvorkin:2020xga}, $m_\chi$ is chosen such that the half-mode (defined as the scale where the DM power spectrum is suppressed by 50\% relative to CDM) matches that of the WDM spectrum. This methodology is supported by what highlighted by Ref.~\cite{Murgia:2018now}. However, as evident in Fig.~3 of Ref.~\cite{Dvorkin:2020xga}, the resulting allowed power spectrum for the non-thermal distribution contains a high-power region that lies below the corresponding WDM spectrum and therefore suppresses power on scales larger than the half-mode. Instead, our criterion for imposing the mass bound on quasi-thermal PSDs follows the criterion established in Ref.~\cite{Konig:2016dzg}. This criterion is further supported by Ref.~\cite{DEramo:2020gpr}, which shows near-perfect agreement between their area-based criterion, a more sophisticated integrated power criterion, and the warmness bound derived from the second moment $\sigma_q$ of the DM PSD. The origin of the discrepancy between these two mass bounds is thus explained: achieving an allowed power spectrum requires a larger high-$k$ cutoff, which in turn implies a higher $m_\chi^{\rm min}$.\footnote{We have explicitly verified that our bound is consistent with results obtained by inputting the PSD into the Boltzmann solver \texttt{CLASS}~\cite{Lesgourgues:2011rh} and computing the power suppression at small scales.}


\section{Conclusions}
\label{sec:conclusions}

The freeze-in mechanism is an appealing framework with many merits, though it remains highly elusive to experimental searches. In this work, we investigated its potential imprints on cosmological observables at small scales. The reason why DM produced via freeze-in can leave such signatures lies in a key feature of light DM with (quasi)-thermal origin: its free-streaming velocities can be large enough to suppress the formation of the smallest structures. Since this suppression depends sensitively on the PSD of freeze-in DM, accurately predicting this distribution through the Boltzmann equation is essential.

\begin{figure*}
\centering
\includegraphics[width=1\linewidth]{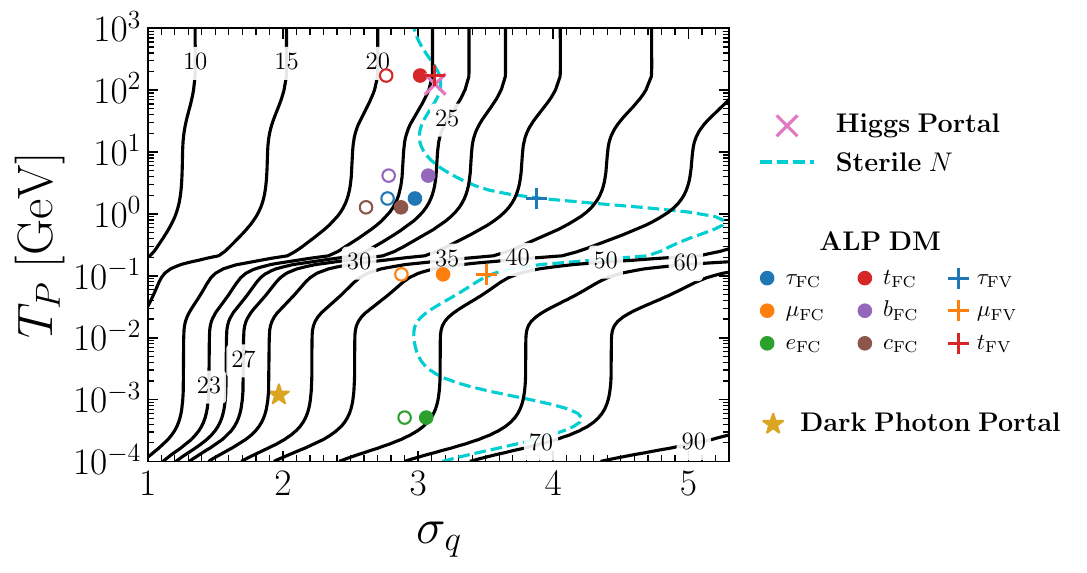}
\caption{This figure summarizes all the results obtained in this paper. The variable on the horizontal axis, $\sigma_q$, is defined in Eq.~\eqref{eq:sigmaq} as the second moment of the PSD evaluated in terms of the comoving momentum. It quantifies the width of the momentum-space distribution. The vertical axis shows the variable $T_P$, which enters the definition of the comoving momentum in Eq.~\eqref{eq:qcom}. While $T_P$ is a conventional parameter, it is convenient to identify it with the temperature at which freeze-in is most efficient—typically corresponding to the mass of the heaviest bath particle participating in DM production. The black isocontours (identical to those in Fig.~\ref{fig:general_bounds}) represent the minimum allowed DM mass produced via freeze-in. These values are obtained through the rescaling given in Eq.~\eqref{eq:warmness_map}, using $m_{\rm WDM}^{\rm min} = 6.8\,\mathrm{keV}$ as a reference. Each model realization is indicated in the plot by the symbols shown in the legend. In each case, $T_P$ is chosen appropriately as the mass of the heaviest particle involved in the decay or scattering process that produces DM. For the dark photon portal, $T_P$ is defined by the condition $\omega_p(T_P) = 2m_\chi$. For the ALP DM scatterings we show the bounds obtained by assuming constant matrix elements as empty circles. We show the sterile neutrino model as a dashed {\bf \color[HTML]{10d1d3} cyan} line, since the mass of the decaying particle is not specified, therefore $T_P$ is also varying; the Higgs portal case is shown by a {\bf \color[HTML]{e997cf} pink} cross. the dark photon portal as a {\bf \color[HTML]{daa521} gold} star. The dots refer to ALP DM produced by flavor conserving (FC) scatterings of SM fermions with gauge bosons, while the pluses to ALP DM produced by flavor violating (FV) decays of SM fermions. 
{\bf \color[HTML]{1f77b4}Blue} markers indicate processes with the tau lepton, {\bf \color[HTML]{ff7f0f
} orange} with the muon, {\bf \color[HTML]{2ba02b
} green} with the electron, {\bf \color[HTML]{d62727
} red} with the top quark, {\bf \color[HTML]{9467bd
} violet} with the bottom and {\bf \color[HTML]{8c564c
} brown} with the charm.}
\label{fig:models_summary}
\end{figure*}

A simple and efficient way to assess whether small-scale structures are preserved is to compare freeze-in DM with a quasi-thermal PSD to the well-studied case of thermal WDM, which serves as a benchmark for the impact of non-zero free-streaming velocities. In this work, we developed a general methodology based on the criterion introduced in Ref.~\cite{Kamada:2019kpe} and numerically validated in Ref.~\cite{DEramo:2020gpr}, where the key mapping variable is the root-mean-square velocity, $W_\chi = \sqrt{\langle{p}^2\rangle}/m_\chi$. This approach allowed us to bypass the need for computationally intensive cosmological Boltzmann solvers such as \texttt{CLASS}~\cite{Lesgourgues:2011rh} when scanning the parameter space of explicit freeze-in models. Thanks to this strategy, we were able to extract lower bounds on the DM mass across a wide range of scenarios.

The plot in Fig.~\ref{fig:models_summary} offers a compact summary of all the results presented in this paper. The two axes show the dimensionless second moment of the PSD, $\sigma_q$ (defined in Eq.~\eqref{eq:sigmaq}), and the temperature scale $T_P$ introduced in Eq.~\eqref{eq:qcom} to define the comoving momentum. Solid black lines indicate the lower bounds on the DM mass obtained via the rescaling in Eq.~\eqref{eq:warmness_map}. For this figure, we adopt the most stringent constraint from MW satellite counts, $m_{\rm WDM}^{\rm min} = 6.8\, {\rm keV}$. As explained in the main text, these bounds are independent of the \textit{arbitrary} choice of $T_P$, which can be selected conveniently. It is natural to set $T_P$ to the largest mass involved in the production process, a choice we adopt for all microscopic realizations analyzed in this work. In the case of the dark photon portal, where both plasmon decay and electron annihilation contribute, $T_P$ is defined as the temperature at which the plasmon frequency equals $2m_\chi$, which exceeds the electron mass. Once $T_P$ is fixed, $\sigma_q$ is correspondingly determined, and each model appears as a single point in the $(\sigma_q, T_P)$ plane. All models discussed in the paper are represented this way, as indicated in the legend. The minimum allowed DM mass can then be directly inferred from the location of the point relative to the black isocontours.

The usefulness of presenting the results in this format lies in the fact that it enables a rapid visualization of different models and provides insight into which freeze-in DM candidates can be considered ``colder'' and which ``warmer.'' For instance, ALPs produced via scatterings with electrons and DM produced through the dark photon portal were both assigned similar values of $T_P$, as production in these scenarios naturally occurs at comparable energy scales. However, the corresponding $\sigma_q$ differ significantly, resulting in DM from the dark photon portal being much colder than that from ALP-electron scatterings. This distinction arises from the different production mechanisms and associated kinematics in the two models. The roles of $T_P$ and $\sigma_q$ are clearly illustrated: $T_P$ serves as a convenient reference scale, typically set by the dominant energy scale of the production process, while $\sigma_q$, once defined with respect to this scale, quantifies the ``warmness'' of the resulting DM population.

Another important result of this work becomes manifest when comparing the model-independent analysis presented in Fig.~\ref{fig:phase-space-suppression} with the summary plot shown in Fig.~\ref{fig:models_summary}. The former was obtained under the assumptions of constant matrix elements for all production processes and Maxwell-Boltzmann statistics for all bath particles. In contrast, for each microscopic realization, we evaluated the transition amplitudes self-consistently and included quantum statistical effects appropriate to the particles involved. For the ALP DM produced via binary scatterings, we show the bounds on the DM mass obtained under the constant matrix element--- that has been employed in setting model-independent bounds in Sec.~\ref{sec:FI}--- as empty circles. Remarkably, the resulting model points align with the model-independent predictions up to 5\%-10\% accuracy, with the constant matrix element assumption providing a more conservative bound. An important exception is the dark photon portal, for which the assumption of a single production mechanism does not hold due to contamination from plasmon decays. This agreement is not unexpected: IR ensures that the momentum dependence of the matrix elements has only a minor impact, and Ref.~\cite{DEramo:2020gpr} explicitly checked that quantum degeneracy effects can be safely neglected for freeze-in production in the early universe. These findings validate the robustness of our model-independent mass bounds under the assumption of no interferences between production mechanisms.

To summarize, this work provided an explicit and computationally efficient methodology to derive mass bounds on freeze-in DM in a model-independent manner. The results presented here lay the foundation for future studies involving DM particles produced with a quasi-thermal momentum distribution. Notable examples include scenarios with a subdominant DM component~\cite{DEramo:2020gpr,Iles:2024zka}, UV-dominated freeze-in production~\cite{Elahi:2014fsa,McDonald:2015ljz,Chen:2017kvz,Bernal:2019mhf,Ballesteros:2020adh,Barman:2022tzk,Ahmed:2022tfm,Ghosh:2023tyz,Freese:2024ogj,deSouza:2024oaz,Caloni:2024olo,Bernal:2025fdr}, or freeze-in from a decoupled thermal bath with a temperature different from that of the SM degrees of freedom~\cite{Fernandez:2021iti}. We leave these directions for future investigation.

\section*{Acknowledgments}

We thank Mathias Becker for useful discussions. F.D. is supported by Istituto Nazionale di Fisica Nucleare (INFN) through the Theoretical Astroparticle Physics (TAsP) project.  This work is supported in part by the Italian MUR Departments of Excellence grant 2023-2027 ``Quantum Frontiers''.  The work of A.L. is supported by an ERC STG grant (``Light-Dark'', grant No. 101040019). This project has received funding from the European Research Council (ERC) under the European Union’s Horizon Europe research and innovation programme (grant agreement No. 101040019).  Views and opinions expressed are however those of the author(s) only and do not necessarily reflect those of the European Union. The European Union cannot be held responsible for them. A.L is grateful to the Azrieli Foundation for the award of an Azrieli Fellowship and to the Department of Physics and Astronomy in Padua for the hospitality during the completion of this work. This article is based upon work from COST Action COSMIC WISPers CA21106, supported by COST (European Cooperation in Science and Technology). AD is supported by the Kavli Institute for Cosmological physics at the University of Chicago through an endowment from the Kavli Foundation and its founder Fred Kavli.

\appendix

\section{The WDM paradigm}
\label{app:WDM}

In this Appendix, we review the WDM framework. The central assumption is that DM is distributed in momentum space according to a thermal PSD. We define $T_{\rm WDM}$ as the current temperature of the WDM population, which may differ from the present CMB temperature $T_0$. The PSD takes the explicit form
\begin{equation}
f_{\rm WDM}(p) = \left\{ 
\begin{array}{lccc}
\left[\exp(p/T_{\rm WDM}) - 1 \right]^{-1} & & & \text{(BE)} \\
\left[\exp(p/T_{\rm WDM}) + 1 \right]^{-1} & & & \text{(FD)} \\
\exp(- p/T_{\rm WDM}) & & & \text{(MB)} 
\end{array}
\right. \ ,
\label{eq:WDM}
\end{equation}
corresponding to Bose-Einstein (BE), Fermi-Dirac (FD), and Maxwell-Boltzmann (MB) statistics. The framework is specified by two parameters: the DM mass $m_{\rm WDM}$ and its temperature $T_{\rm WDM}$, which determines the shape of the distribution independently of the thermal bath.

For a WDM candidate with $g_{\rm WDM}$ internal degrees of freedom (e.g., $g_{\rm WDM} = 2$ for a Majorana fermion), the DM number density is obtained by integrating the phase-space distribution
\begin{equation}
\begin{split}
n_{\rm WDM}(t_0) & = g_{\rm WDM} \int \frac{d^3 p}{(2\pi)^3} f_{\rm WDM}(p) \\ 
& = g_{\rm WDM} \, g_{\rm eff} \frac{\zeta(3)}{\pi^2} T_{\rm WDM}^3 \,,
\end{split}
\end{equation}
where $\zeta(3)$ is the Riemann zeta function. The statistical factor $g_{\rm eff} = \{ 1,\ 3/4,\ 1/\zeta(3) \}$ arises from the momentum integral of the PSD in the BE, FD, and MB cases, respectively, as listed in Eq.~\eqref{eq:WDM}.

The current WDM energy density is given by multiplying the number density at the present time $t_0$ by the WDM mass, $\rho_{\rm WDM}(t_0) = m_{\rm WDM} \, n_{\rm WDM}(t_0)$. It is convenient to express this quantity in terms of the usual parameter $\Omega_{\rm WDM}$, defined as
\begin{equation}
\Omega_{\rm WDM} h^2 \equiv \frac{\rho_{\rm WDM}(t_0)}{\rho_{\rm cr}(t_0)/h^2} \,,
\end{equation}
where $\rho_{\rm cr}(t_0) \simeq 1.05 \times 10^{-5} \, h^2 \, {\rm GeV} \, {\rm cm}^{-3}$ is the current critical density~\cite{ParticleDataGroup:2024cfk}. Combining the above expressions, the WDM temperature can be related to the present CMB temperature $T_0$ as follows
\begin{equation}
\frac{T_{\rm WDM}}{T_0} \simeq 0.16 \left( \frac{2}{g_{\rm WDM}} \, \frac{3/4}{g_{\rm eff}} \, \frac{\rm keV}{m_{\rm WDM}} \, \frac{\Omega_{\rm WDM} h^2}{0.12} \right)^{1/3} \,.
\end{equation}
Here, we have normalized the expression using parameters appropriate for a Majorana fermion, as WDM mass bounds are typically quoted under this assumption.

At this point, we can evaluate the quantity defined in Eq.~\eqref{eq:Wchi}, which characterizes the WDM warmness
\begin{equation}
W_{\rm WDM} = \frac{\sqrt{\langle p^2 \rangle}}{m_{\rm WDM}} = \sigma_q^{\rm WDM} \, \frac{T_{\rm WDM}}{m_{\rm WDM}} \,,
\end{equation}
where $\sigma_q^{\rm WDM}$ is the dimensionless second moment of the WDM distribution, given by
\begin{equation}
\sigma_q^{\rm WDM} = \left\{ 
\begin{array}{lccc}
2 \sqrt{3 \, \zeta(5) / \zeta(3)} \simeq 3.22 & & & \text{(BE)} \\
\sqrt{15 \, \zeta(5) / \zeta(3)} \simeq 3.60 & & & \text{(FD)} \\
2 \sqrt{3} \simeq 3.46 & & & \text{(MB)} 
\end{array}
\right. \,.
\end{equation}

\section{Collision operators and analytic PSDs}
\label{app:Cs}
We summarize the collision operators for the topologies used in this work and the analytical expressions for the PSD. These and other related results can be found in Ref.~\cite{DEramo:2020gpr} with the exception of the new analytical PSD for three-body decays given in Eq.~\eqref{eq:PSD3}. We work in the light DM limit, $m_\chi \ll m_1$, and we report the expressions for DM single production. As discussed in the text, these results are valid also for multiple production unless we have almost degenerate bath particles. The analytical expressions of the PSD provided here match the numerical results to high accuracy in the limit $g_* = {\rm const}$.

\subsection{Two-body decays}
For a two-body decay the collision term reads
\begin{align}
    C_{1\to 2\chi} = \frac{\ell}{2}\int d\Pi_1 d\Pi_2 (2\pi)^4 \delta^{(4)}(P_1 -P_2 - P)|{\cal M}|^2 f_1^{\rm eq}\ .
\end{align}
The matrix element is always constant, and the phase space integration leads to the following result
\begin{align}
    C_{1\to 2\chi}(p,T) = \frac{\ell g_1 g_2 T}{16\pi p} |{\cal M}|^2    h(p,T)\ .
\end{align}
The dimensionless function $h(p,T)$ depends on the statistics of the decaying bath particle
\begin{align}
    h(p,T) \approx \begin{cases}
        e^{-(E_2^-+E)/T}-e^{-(E_2^++E)/T} & {\rm (MB)} \\
       - \log \left[\dfrac{1- e^{-(E_2^-+E)/T}}{ 1- e^{-(E_2^++E)/T}}\right] & {\rm (BE)}\\ 
       \log \left[\dfrac{1+ e^{-(E_2^-+E)/T}}{ 1+ e^{-(E_2^++E)/T}}\right] & {\rm (FD)} 
    \end{cases}
\end{align}
and the bounds on $E_2$ being $E_2^+ \to \infty$
\begin{align}
    E_2^- \approx  \frac{m_1^2}{4p T}\left( 1- \frac{m_2^2}{m_1^2}\right) +\frac{ pm_2^2}{m_1^2-m_2^2} + {\cal O}(m_\chi)\, .
\end{align}
An analytical PSD estimate can be obtained by integrating the collision term assuming $g_* = {\rm const}$, radiation domination, MB statistics, and $q=p/T$. We find
\begin{align}
    f_\chi(q) \propto  \frac{1}{\sqrt{q}} e^{-q (1-m_2^2/m_1^2)^{-1}}\ .
\end{align}
This implies $\sigma_q \simeq  2.958\sqrt{1-(m_2/m_1)^2}$. Note how the phase space suppression $m_2/m_1$ makes the PSD ``colder'', reducing the momentum dispersion. 

\subsection{Binary scatterings}
For a binary scattering we have
\begin{align}
    C_{12\to 3\chi} &= \frac{\ell}{2}\int \prod_{i=1}^3 d\Pi_i (2\pi)^4 \delta^{(4)}(P_{\rm i} -P_{\rm f}) 
   |{\cal M}|^2 f_1^{\rm eq} f_2^{\rm eq}\ .
\end{align}
The expression greatly simplifies for MB statistics
\begin{align}
    C_{12\to 3\chi} = &\frac{\ell g_1 g_2 g_3}{256\pi^3}\frac{T e^{-E/T}}{p}\nonumber \\
    &\quad \times \int_{s_-}^{\infty} \frac{ds\, e^{-E_3^-/T}}{\sqrt{\lambda (s,m_3,m_\chi)}} \int _{t_-}^{t_+} dt\, |{\cal M}|^2 \ ,
\end{align}
we have introduced the Källén function $\lambda(a,b,c) \equiv [a-(b+c)^2][a-(b-c)^2]$. Here $E^-_3 \approx \sqrt{m_3^2 + (s/4p)^2}+ {\cal O}(m_\chi)$ and the integration limits read
\begin{align}
    s_- &= \max[(m_1+m_2)^2 , (m_3+m_\chi)^2]\\
    t_{\pm}& =m_2^2 + m_\chi^2 - 2\left( \sqrt{m_2^2 +p_{12}^2}\sqrt{m_3^2 +p_{3\chi}^2} \mp p_{12}p_{3\chi} \right) 
\end{align}
where $p_{ij}(s) = \sqrt{\lambda(s,m_i,m_j)/s}/2$.

The matrix element is not constant in general for this case. If we approximate it to a constant value, and with the further assumptions of constant $g_*$, radiation domination, MB statistics, $q=p/T$, and $m_2 = m_3=0$, we find the analytical solution
\begin{align}
    f_\chi(q) &\propto \int \frac{dT \, e^{-p/T}}{T^2 Hp^2} \int_{m_1^2}^\infty ds \frac{s-1}{s}e^{-{s}/{4pT}} \propto \frac{1}{\sqrt{q}}e^{-q} \ .
\end{align}
This implies $\sigma_q \simeq 2.958 $.

\subsection{Three-body decays}
Finally, for three-body decays we have 
\begin{align}
  C_{1 \to 23\chi} &= \frac{\ell}{2}\int \prod_{i=1}^3 d\Pi_i (2\pi)^4 \delta^{(4)}(P_{\rm i} -P_{\rm f}) 
   |{\cal M}|^2 f_1^{\rm eq} \ .
\end{align}
Also in this case the matrix element is not constant and has to be integrated over specially defined Mandelstam-like variables $s=(P_1-P)^2$ and $t=(P_1-P_2)^2$. If we assume MB statistics
\begin{align}
    C_{1\to 23\chi} = &\frac{\ell g_1 g_2 g_3}{256\pi^3}\frac{T}{p}\nonumber \\
    &  \times \int_{s_-}^{s_+} \frac{ds\, e^{-E_1^-/T}}{\sqrt{\lambda (s,m_1,m_\chi)}} \int _{t_-}^{t_+} dt\, |{\cal M}|^2\ . 
\end{align}
The integration limits read
\begin{align}
    s_+ &= (m_1-m_\chi)^2,  
    \qquad s_-  =  (m_2+m_3)^2,  \\
    t_{\pm}& =m_3^2 + m_\chi^2 + 2\left( \sqrt{m_3^2 +p_{23}^2}\sqrt{m_\chi^2 +p_{1\chi}^2} \pm p_{1\chi}p_{23} \right) ,
\end{align}
where $p_{ij}(s) = \sqrt{\lambda(s,m_i,m_j)/s}/2$.  $E_1^-$ is given as
\begin{align}
     E^-_1 \approx    \frac{p}{1-s/m_1^2} + \frac{m_1^2-s}{4p}    + {\cal O}(m_\chi) \ .
\end{align}

We can obtain (a rather peculiar) analytical form for the PSD. The assumptions are again constant $g_*$, radiation domination, MB statistics, $|{\cal M}|^2$  constant, $q=p/T$, $m_3=0$. We keep $m_2$ non-zero to show the phase space suppression, and we find
\begin{align}\nonumber
    f_\chi(q)&\propto \int \frac{dT}{H} \frac{1}{p^2}\int_{{m_2^2}}^{m_1^2} ds\    e^{-E_1^- /T}\left (1-\frac{m_2^2}{ s}\right)\\ 
    &\propto  \frac{1}{q}{\rm erfc} \left(\sqrt{\frac{q}{1-m_2^2/m_1^2}}\right)
\label{eq:PSD3}
\end{align}
In this case the integral over temperatures was performed before the one over $s$. Remarkably, we obtained an expression that takes into account the phase space suppression. We obtain $\sigma_q \simeq  2.092 \sqrt{1-(m_2/m_1)^2}$, showing how the mass of particle 2 makes the PSD colder. Note also how, even for $m_2=0$, three-body decays have a smaller $\sigma_q$ than two-body decays.

\bibliographystyle{apsrev4-2}
\bibliography{ref}

\end{document}